\documentclass[10pt]{iopart}

\usepackage{bbm,mathrsfs}
\usepackage{graphicx}
\usepackage{amsfonts}

\usepackage{iopams}

\newcommand\T{\rule{0pt}{5.6ex}}
\newcommand\TT{\rule{0pt}{4.6ex}}
\newcommand\TTT{\rule{0pt}{3.0ex}}
\newcommand\B{\rule[-3.2ex]{0pt}{0pt}}
\newcommand\BB{\rule[-2.0ex]{0pt}{0pt}}

\def\textbf#1{{\bf #1}}
\def\be{\begin{equation}}
\def\ee{\end{equation}}
\def\ben{\begin{eqnarray}}
\def\een{\end{eqnarray}}
\def\eea{\end{array}}
\def\bea{\begin{array}}
\newcommand{\ot}[0]{\otimes}
\newcommand{\bei}{\begin{itemize}}
\newcommand{\eei}{\end{itemize}}
\newcommand{\ket}[1]{|#1\rangle}
\newcommand{\bra}[1]{\langle#1|}

\newcommand{\proj}[1]{\ket{#1}\!\bra{#1}}

\def\pcal{{\cal P}}

\begin{document}

\title[]{Positive maps, majorization, entropic inequalities, and detection of entanglement}

\author{R Augusiak}

\address{ICFO--Institute Ci\`{e}ncies Fot\`{o}niques,
Mediterranean Technology Park, 08860 Castelldefels (Barcelona),
Spain} \ead{remigiusz.augusiak@icfo.es}

\author{J Stasi\'nska}

\address{Grup de F\'isica Te\`orica, Universitat Aut\`onoma de
Barcelona, 08193 Bellaterra (Barcelona), Spain}

\ead{julsta@gmail.com}

\begin{abstract}

In this paper, we discuss some general connections between the 
notions of positive map, weak majorization and entropic inequalities in the
context of detection of entanglement among bipartite quantum systems. First, 
basing on the fact that any positive map $\Lambda:M_{d}(\mathbb{C})\to M_{d}(\mathbb{C})$ can be written 
as the difference between two completely positive maps $\Lambda=\Lambda_{1}-\Lambda_{2}$, we 
propose a possible way to generalize the Nielsen--Kempe majorization criterion. 
Then we present two methods of derivation of some general classes of 
entropic inequalities useful for the detection of entanglement. While the first
one follows from the aforementioned generalized majorization relation and the 
concept of the Schur--concave decreasing functions, the second is based on some 
functional inequalities. What is important is that, contrary to the Nielsen--Kempe 
majorization criterion and entropic inequalities, our criteria allow for the 
detection of entangled states with positive partial transposition when using 
indecomposable positive maps. We also point out that if a state with at least 
one maximally mixed subsystem is detected by some necessary criterion based 
on the positive map $\Lambda$, then there exist entropic inequalities derived from $\Lambda$ (by
both procedures) that also detect this state. In this sense, they are equivalent to the 
necessary criterion $[I\ot\Lambda](\varrho_{AB})\geq 0$. Moreover, our inequalities provide a way 
of constructing multi--copy entanglement witnesses and therefore are promising 
from the experimental point of view. Finally, we discuss some of the derived 
inequalities in the context of recently introduced protocol of state merging 
and possibility of approximating the mean value of a linear entanglement 
witness.

\end{abstract}
\maketitle

\section{Introduction}

Due to its interesting applicability, entanglement (see e.g. \cite{przegladowka}) is still one of the most 
interesting topics in modern physics. Following \cite{Werner}, we call a state describing some finite--dimensional bipartite physical system $A$ and $B$ {\it entangled} if it cannot be written as a mixture 
of product states. More precisely, $\varrho_{AB}$ is called {\it entangled} if it does not admit the following decomposition:
\begin{equation}\label{separable}
\varrho_{AB}=\sum_{i}p_{i}\varrho_{A}^{(i)}\ot \varrho_{B}^{(i)},
\qquad p_{i}\geq 0, \qquad \sum_{i}p_{i}=1,
\end{equation}
where $\varrho_{A(B)}^{(i)}$ are density matrices representing subsystems $A$ and $B$. In the case when $\varrho_{AB}$ can 
be written in the above form, it is called {\it separable} or {\it classically correlated}.

One of the most important problems the scientists have to face is that there is no simple way 
of deciding if a given state $\varrho_{AB}$ is entangled or not. The general problem of separability remains 
unresolved despite the fact that huge effort has been expended so far to invent stronger and 
easier to apply separability criteria (see e.g. \cite{Peres,HHH_entr,mapy,range,NielsenKempe,Korbicz1,Korbicz,Guhne,deVicente,ChengJieZhang,covariance,SperlingVogel,SpinSq} and the recent review on the detection of 
entanglement \cite{EntDet}). The task becomes even harder since most of the invented mathematical 
criteria are not directly applicable experimentally. In general, despite the case of pure states 
(see \cite{pure}) and arbitrary two--qubit states (see \cite{RAMDPH}), there is no unambiguous method for deciding 
about separability that is promising from the experimental point of view. Therefore, it is still 
desirable to look for separability criteria giving, on the one hand, mathematical possibility 
of detecting entanglement and, on the other hand, allowing for future experimental realizations.

Still, one of the most significant separability criteria, but in general not directly realizable in
experiment, are those based on positive maps. It was pointed out first for the partial transposition 
in \cite{Peres} and then generally in \cite{mapy} that a positive map can serve as a necessary separability criterion. That is, whenever $\varrho_{AB}$ is separable, the operator inequality
\begin{equation}\label{operatorInequality}
[I\ot\Lambda](\varrho_{AB})\geq 0
\end{equation}
holds for all positive but not completely positive maps $\Lambda$. Here, by $I$ we denoted an identity 
map acting on the first subsystem. In fact, even a stronger statement is true \cite{mapy}, i.e. $\varrho_{AB}$ 
is separable if and only if condition (\ref{operatorInequality}) is satisfied for all positive maps $\Lambda$. Exemplary maps 
serving commonly for the detection of entanglement are the reduction map \cite{reduction,CerfAdamiGingrich} $\Lambda_{R}=\Lambda_{\mathrm{Tr}}-I$ 
with $\Lambda_{\mathrm{Tr}}(X)=\Tr(X)\mathbbm{1}_{d}$ and the transposition map (hereafter denoted by $T$) \cite{Peres,mapy}.
Interestingly, it was shown in \cite{mapy} that in the case of the $2\otimes 2$ and $2\otimes 3$ system the transposition 
map completely solves the separability problem, as in this case $[I\ot T](\varrho_{AB})\equiv\varrho^{\Gamma}_{AB}\geq 0$ is a 
necessary and sufficient condition for separability. Though the criteria cannot be applied directly 
in experiment, some indirect methods were pointed out in a series of papers \cite{maps_exp}.

On the other hand, there also exist criteria that seem to be promising from the experimental 
point of view and can be given clear physical interpretation. Among others are the so--called
entropic inequalities. Their meaning for detection of entanglement was first pointed out
in \cite{RHPH}, where it was shown that for separable states the von Neumann conditional entropy 
satisfies $S(B|A;\varrho_{AB})\equiv S(\varrho_{AB})- S(\varrho_{A(B)})\geq 0$, where $S$ denotes the von Neumann entropy
$S(\sigma)=-\Tr\sigma\log\sigma$. This means that in the case of separable states the whole system is more 
disordered than its subsystems. However, the nonnegativity of conditional entropy does not 
have to hold for all quantum states (as it is easily seen in the case of pure entangled states). This 
phenomenon found its explanation in \cite{MHJOAW}, where the conditional von Neumann entropy was 
interpreted as the cost of merging of a bipartite state. The fact that for some entangled states this 
quantity can be negative is one of the basic features of quantum communication.

Further, in a series of papers \cite{HHH_entr,RHMH,Terhal,VollbrechtWolf}, the property that for separable states the whole system is more disordered than the subsystems was formalized in terms of other disorder 
measures, for instance the quantum version of the Renyi entropy \cite{Renyi} (or equivalently the Tsallis 
entropy \cite{Tsallis}). This led to inequalities of the form
\begin{equation}\label{entr}
S_{\alpha}^{R}(\varrho_{A(B)})\leq S_{\alpha}^{R}(\varrho_{AB})
\qquad (\alpha\geq 0)
\end{equation}
with $S^{R}_\alpha(\sigma)=[1/(1-\alpha)]\log\Tr\sigma^{\alpha}$ (see e.g. \cite{Wehrl}). Applicability in the detection of 
entanglement among many particular examples of quantum states were investigated in the 
literature (see e.g. \cite{Investigation,ZHHH}). Also, other entropic functions were studied in this context \cite{RossignoliCanosa}.

This disorder rule was also connected to a more general disorder measure based on the 
concept of majorization (the definition of majorization will be given in Section 2). In \cite{NielsenKempe}, it was shown that if $\varrho_{AB}$ is separable, then the following holds:
\begin{equation}\label{majoryzacja}
\lambda(\varrho_{A})\succ\lambda(\varrho_{AB})\quad \wedge\quad
\lambda(\varrho_{B})\succ\lambda(\varrho_{AB}),
\end{equation}
where $\lambda(\varrho_{AB})$ and $\lambda(\varrho_{A(B)})$ denote vectors consisting of eigenvalues of $\varrho_{AB}$ and $\varrho_{A(B)}$, 
respectively (note that one has to add zeros to $\lambda(\varrho_{A(B)})$ to have the same dimension as $\lambda(\varrho_{AB})$).

The advantage of the entropic inequalities lies in the fact that they can be measured 
experimentally. More precisely, since they can be rewritten as
\begin{equation}
\label{entrWithout1}\Tr\varrho_{A(B)}^{\alpha}\geq
\Tr\varrho^{\alpha}_{AB}\qquad (\alpha\in[1,\infty))
\end{equation}
\begin{equation}
\label{entrWithout2}\Tr\varrho_{A(B)}^{\alpha}\leq
\Tr\varrho^{\alpha}_{AB}\qquad (\alpha\in[0,1)),
\end{equation}
they lead, for integer $\alpha$, to multicopy entanglement witnesses on $\alpha$ copies of $\varrho_{AB}$
\footnote{Following \cite{manycopy} we say that a Hermitian operator $W^{(n)}$ is $n$--copy entanglement witness if its mean value on $n$ 
copies of any separable state is positive and is negative on at least one entangled state.}
\cite{manycopy}. For 
$\alpha=2$, this fact has already been already applied experimentally \cite{Bovino,Schmid} and a possible generalization to $n$--qubit states was considered in \cite{Jaksch}.

On the other hand, both criteria (majorization relations and entropic inequalities) were 
shown to follow from the reduction criterion \cite{VollbrechtWolf,Hiroshima}. This means that they are useless in
detecting entangled states with positive partial transposition (PPT) \cite{BE}, since the reduction map
is decomposable, and in general weak. In a series of two papers \cite{RAJSPH,RAJS}, the question about other 
entropic or entropic--like inequalities that could detect bound entanglement and could be realized 
experimentally was posed. The fact that the reduction criterion map leads to some scalar criteria
suggests that it should be also possible to get some entropic--like inequalities from other maps, 
including indecomposable ones. Using the fact that any positive map $\Lambda:M_{d}(\mathbb{C})\to M_{d}(\mathbb{C})$ may be written as $\Lambda=\Lambda_{1}-\Lambda_{2}$ with $\Lambda_{i}$ ($i=1,2$) being two completely positive maps, the 
authors derived in \cite{RAJS} a class of inequalities efficiently detecting entanglement. However, these 
inequalities can be applied only to states fulfilling some commutation relations.

The purpose of the present paper is to reexamine the question about the possibility of 
deriving entropic--like inequalities from any positive map, not only the reduction one, but
without any additional assumptions on the state. Analogously, we ask if there are some 
majorization criteria following from any positive map. This in turn, by the concept of the Schur--convex 
functions, could give a general method of deriving scalar inequalities appropriate for 
experimental realization. In what follows we show that it is possible to get submajorization 
relations following from positive maps. Then, we provide two methods allowing for the 
derivation of entropic inequalities. The first one is a simple consequence of the submajorization 
relations, while the second one is a continuation of the results from \cite{RAJS}. Both constructions give 
separability tests allowing for detection of PPT entangled states. Interestingly, for states with 
one of the subsystems being in a maximally mixed state the derived inequalities are in some 
sense equivalent to the necessary condition based on the positive map from which they were 
derived. Finally, the particular inequalities derived using the second method can be given some 
physical meaning in the context of state merging and approximation of a mean value of a 
linear entanglement witness.

The paper is organized as follows: Section \ref{whatweknow} contains various definitions concerning majorization, 
week majorization, positive maps and related concepts. The reader familiar with the subject 
matter can move straight to Section \ref{inequalities}, where the inequalities are presented together with some 
special cases and comments. In Section \ref{comparison}, we analyze the effectiveness of the inequalities as a 
separability criterion. A summary of obtained results and some open questions are given in 
section \ref{conclusion_sec}.

\section{Definitions}\label{whatweknow}

\subsection{Majorization and the Schur--concave functions}

In what follows we shall often refer to the notion of majorization. It is therefore desirable 
to recall the definition and some concepts closely related to it. The meaning of majorization 
in physics has already been recognized (see e.g. \cite{AlbertiUhlmann} and references therein). Recently, 
majorization found its applications also in issues closely related to quantum information theory 
(see e.g. \cite{NielsenKempe,NielsenConditions,NielsenVidal}).

All the concepts appearing in this subsection can be found e.g. in \cite{Bhatia,Marshall}. Let then $x$ 
be a vector from $\mathbb{R}^{d}$ and $x^{\downarrow}=(x_{0}^{\downarrow},\ldots,x^{\downarrow}_{d-1})\in\mathbb{R}^{d}$ denote a vector consisting of coordinates of $x$ 
put in decreasing order, i.e. $x_{0}^{\downarrow}\geq x_{1}^{\downarrow}\geq\ldots\geq x_{d-1}^{\downarrow}$. We say that $x$ majorizes $y$, which we write 
as $x\succ y$, if the conditions
\begin{equation}
\sum_{i=0}^{k}x_{i}^{\downarrow}\geq
\sum_{i=0}^{k}y^{\downarrow}_{i}
\end{equation}
hold for any $k=0,\ldots,d-2$ and
\begin{equation}
\sum_{i=0}^{d-1}x_{i}^{\downarrow}=
\sum_{i=0}^{d-1}y^{\downarrow}_{i}.
\end{equation}
However, if in the last condition one replaces equality with inequality `$\geq$', then we say that 
$x$ {\it weakly majorizes } (or {\it submajorizes}) $y$, which we shall denote by $x \,\,{}_{w}\!\!\succ y$.

The notions of majorization and weak majorization are strictly connected to the notion 
of doubly stochastic and doubly substochastic matrices. Let then $S$ denote some $d\times d$ matrix. 
We say that $S=(S_{ij})$ is a {\it doubly stochastic} ({\it substochastic}) matrix if $S_{ij}\geq 0$ and the sum of
all elements in any row and column equals one (is less than or equal to one). Using the above 
notions, we can reformulate the concept of majorization as follows (see e.g. \cite{Bhatia}).\\

\noindent{\it {\bf Fact 1.} Let $x,y\in\mathbb{R}^{d}$. Then $x\succ y$ ($x \,\,{}_{w}\!\!\succ y$) if and only if there exists such a doubly stochastic 
(doubly substochastic) matrix $S$ that
\begin{equation}
y=Sx.
\end{equation}
}

Another notion useful for further considerations and strictly connected with majorization 
is the Schur--convex function. We say that a real--valued function $\phi:\mathcal{I}^{d}\to\mathbb{R}$ ($\mathcal{I}\subseteq\mathbb{R}$ denotes 
some open interval) is {\it Schur--convex} if the following implication holds:
\begin{equation}
x\succ y\quad\Rightarrow\quad \phi(x)\geq \phi(y).
\end{equation}
If the inequality in the above is reversed, we say that $\phi$ is {\it Schur--concave}. There is an easy--to--apply
tool allowing us to decide if a given function $\phi$ is Schur--convex (concave). Namely,
$\phi:\mathcal{I}^{d}\to\mathbb{R}$ is Schur--convex if and only if it is symmetric and fullfils the following condition:
\begin{equation}\label{conditionSchur}
(x_i-x_j)\left(\frac{\partial \phi(x)}{\partial
x_i}-\frac{\partial \phi(x)}{\partial x_j}\right)\geq 0
\end{equation}
for any $i\neq j$ (see e.g. \cite{Marshall}). Once again, if the inequality in the above is reversed, the function
$\phi$ is {\it Schur--concave}. (Notice that if $\phi$ is symmetric one does not have to check the above condition 
for all $i$ and $j$. It suffices to check it for some pair $i\neq j$.)

The last definition we give in this subsection deals with the monotonicity of a multi--variable 
function. We say that a $d$--argument function $\phi$ is {\it increasing (decreasing)} if for $x,y\in\mathbb{R}^{d}$ and 
$x\geq  y$ one has $\phi(x)\geq\phi(y)$ ($\phi(x)\leq \phi(y)$). Here the ordering relation $x\geq  y$ means that $x_{i}\geq y_{i}$ 
for any $i=0,\ldots,d-1$. If the above inequalities are strict we call a function {\it strictly increasing
(decreasing)}.

\subsection{Positive maps}

Finally, we introduce the concepts of positive and completely positive map and discuss 
some of their properties. Let $\Lambda:M_{d}(\mathbb{C})\to M_{d}(\mathbb{C})$ denote a linear map. We say that $\Lambda$ is 
{\it positive} if, acting on a positive element of $M_{d}(\mathbb{C})$, it returns a positive matrix. We say that 
$\Lambda$ is completely positive if the extended map $I_{k}\ot\Lambda$ is positive for any natural $k$. Here by 
$I_{k}:M_{k}(\mathbb{C})\to M_{k}(\mathbb{C})$ we denoted the identity map. A more detailed characterization of positive 
and completely positive maps can be found, e.g. in \cite{Jam,Choi_iso}. As already stated, any positive 
but not completely positive map constitutes a necessary separability criterion for finite bipartite 
quantum systems. Moreover, we say that a positive map $\Lambda$ is {\it decomposable (indecomposable)} if 
it can (cannot) be written as a sum of a completely positive map and a completely positive map 
composed with transposition map. What is important here is that separability criteria based 
on decomposable positive maps cannot detect PPT entangled states. Only indecomposable maps 
have this advantage.

On the other hand, any positive map on finite--dimensional matrix algebra can be written as 
the difference of two completely positive maps $\Lambda_{1}$ and $\Lambda_{2}$ (see e.g. \cite{Hou}), i.e.
\begin{equation}\label{relation5}
\Lambda=\Lambda_{1}-\Lambda_{2}.
\end{equation}
Then the necessary condition for separability of $\varrho_{AB}$ based on some particular positive map $\Lambda$ reads:
\begin{equation}\label{operatorIn}
[I\ot\Lambda_{1}](\varrho_{AB})\geq
[I\ot\Lambda_{2}](\varrho_{AB}).
\end{equation}
Finally, it should be also reminded that any completely positive map $\Lambda_{CP}:M_{d}(\mathbb{C})\to M_{d}(\mathbb{C})$ can 
be written as \cite{Choi_iso, postacKrausa}
\begin{equation}\label{KrausChoi}
\Lambda_{CP}(\cdot)=\sum_{i=1}^{\kappa}V_{i}(\cdot)V_{i}^{\dagger},
\end{equation}
where $V_{i}$ are some linearly independent $d\times d$ matrices. The parameter $\kappa$ corresponding to the 
smallest number of operators $V_{i}$ in (\ref{KrausChoi}) is called a {\it minimal length} of $\Lambda$ \cite{Jamiolkowski1}. In this way any 
positive map on a finite matrix algebra is characterized by two numbers $\kappa_{1}$ and $\kappa_{2}$ corresponding 
to completely positive maps $\Lambda_1$ and $\Lambda_2$ in its decomposition. The decomposition corresponding 
to minimal $\kappa_1+\kappa_2$ will be called a minimal decomposition. As we shall see later, the efficiency 
of our inequalities in detecting entanglement strongly depends on the choice of $\Lambda_1$ and $\Lambda_2$ in equation (\ref{relation5}).

For simplicity, in further considerations, we shall be denoting an extended map $I\ot\Lambda_{i}$ by $\Theta_{i}$ $(i=1,2)$ and $\varrho_{AB}$ by $\varrho$ whenever it does not lead to confusion.

\section{Inequalities}\label{inequalities}

Now we are ready to provide two constructions of entropic--like inequalities. The first method 
goes through the weak majorization relations generalizing the famous majorization criterion 
by Nielsen and Kempe \cite{NielsenKempe}, while the second one is direct in the sense that beginning with an 
operator inequality (\ref{operatorIn}), we prove some functional inequality which in case of reduction 
map reproduces the standard entropic criterion. Before that, in the following subsection, we will
briefly recall some of the results of \cite{RAJSPH,RAJS}.

\subsection{Already known entropic--like inequalities going beyond the standard ones}
As mentioned previously the standard entropic inequalities, though they possess a physical 
interpretation, are unable to detect PPT entangled states. This is because they follow from the 
reduction criterion \cite{VollbrechtWolf}, which is in general weaker than the criterion based on transposition. 
The question about the possibility of constructing some stronger inequalities resembling the 
entropic ones, using other positive maps (also indecomposable), has been recently addressed 
in \cite{RAJSPH,RAJS}. Firstly, in \cite{RAJSPH} it was shown that using the Breuer--Hall map \cite{Breuer,Hall}:
\begin{equation}
\Lambda_{BH}^{U}(X)=\Tr(X)\mathbbm{1}_{d}-X-\tau^{U}(X) \qquad
(X\in M_{d}(\mathbb{C})),
\end{equation}
where $\tau^{U}(X)=UX^{T}U^{\dagger}$ with $U$ denoting some antisymmetric matrix $(U^{T}=-U)$ obeying 
$U^{\dagger}U\leq \mathbbm{1}_{d}$, one can derive inequalities efficiently detecting PPT states. However, the 
inequalities work only when some commutation relations are satisfied by the investigated state. 
For instance, assuming that $\varrho$ is separable and that the commutation relation $[\varrho,\varrho_{A}\ot\mathbbm{1}_{d}]=0$ 
holds, the following inequality was derived in \cite{RAJSPH}:
\begin{equation}
\Tr\varrho_{A}^{\alpha}\geq
\frac{1}{2}\left\{\Tr\left[\varrho(\varrho+\tau_{B}^{U}(\varrho))^{\alpha-1}\right]+
\Tr\left[\varrho(\varrho-\tau_{B}^{U}(\varrho))^{\alpha-1}\right]\right\}
\end{equation}
with $\alpha\geq 1$. Assuming further that $[\varrho,\tau_{B}^{U}(\varrho)]=0$, we can obtain from the above
\begin{equation}\label{Fact1}
\Tr\varrho_{A}^{\alpha}\geq
\Tr\varrho^{\alpha}+\sum_{k=1}^{\lfloor (\alpha-1)/2\rfloor}
\left(\begin{array}{c}
\alpha-1\\
2k
\end{array}\right)\Tr\left\{\varrho^{\alpha-2k}[\tau_{B}^{U}(\varrho)]^{2k}\right\}.
\end{equation}
In both inequalities $\varrho_{A}=\Tr_{B}\varrho$ and $\tau_{B}^{U}$ denotes the map $\tau^{U}$ applied to the second subsystem.
Comparing the above to the standard entropic inequalities (\ref{entrWithout1}), it is clear that they are stronger,
since on the right--hand hand side we have an additional nonnegative term. Moreover, it was pointed 
out in \cite{RAJSPH} that such inequalities can detect PPT entangled states in some physical systems.

In \cite{RAJS}, the authors posed a question about some more general inequalities following from 
any positive map $\Lambda$. For instance, the following inequalities:
\begin{equation}\label{th1}
\Tr\varrho^{\alpha}\left([I\ot\Lambda_{1}](\varrho)\right)^{\beta}\geq
\Tr\varrho^{\alpha}\left([I\ot\Lambda_{2}](\varrho)\right)^{\beta}\qquad
(\beta\geq 0)
\end{equation}
and
\begin{equation}\label{th3}
\Tr\varrho^{\alpha}\left([I\ot\Lambda_{1}](\varrho)\right)^{\beta}\leq
\Tr\varrho^{\alpha}\left([I\ot\Lambda_{2}](\varrho)\right)^{\beta}
\qquad (\beta\in[-1,0)),
\end{equation}
were derived, both for $\alpha\geq 0$. It should also be explicitly stated than in the case of the second 
inequality (\ref{th3}) one has to remember that both operators $[(I\ot\Lambda_{i})(\varrho)]^{\beta}$ have to be `sandwiched' 
with the projector onto the support of $[I\ot \Lambda_{2}](\varrho)$ (this can always be done during the derivation 
of this inequality, cf proof of Theorem 2) to avoid the problems of inverse of matrices that 
are not of full rank. This means that on the right--hand side one can take the pseudoinverse of 
$[I\ot\Lambda_{2}](\varrho)$.

Unfortunately, to prove both inequalities (\ref{th1}) and (\ref{th3}) (except for the case $\beta\in[0,1)$), one 
has to assume that $[[I\ot\Lambda_{2}](\varrho),\varrho]=0$; however, as discussed in \cite{RAJS}, for many of the known
positive (even indecomposable) maps, $\Lambda_{2}$ can be chosen to be an identity map. This means that 
in these cases the problem of commutativity vanishes. Also, inequalities (\ref{th1}) can detect PPT 
entanglement if derived from indecomposable maps.

Our main aim now is to discuss the possibility of deriving entropic inequalities which do 
not require any additional assumptions, and on the other hand, allow for efficient detection of
PPT entangled states.

\subsection{Weak majorization and entropic inequalities}

We start by proving a simple fact relating the operator inequality $[I\ot \Lambda_{1}](\varrho)\geq [I\ot\Lambda_{2}](\varrho)$ 
and the concept of weak majorization. For this purpose let us prove a kind of generalization of results from \cite{Hiroshima}.\\

\noindent {\it {\bf Theorem 1.} If $A$ and $B$ are such positive operators that $A\geq B$, then
\begin{equation}\label{majorrel}
\lambda(A)\;{}_{w}\!\!\succ\lambda(B).
\end{equation}
}

\noindent{\it Proof.} The proof simply follows the one given in \cite{Hiroshima}. Firstly, let us note that from the
assumption it follows that $\mathrm{ker}(A)\subseteq\mathrm{ker}(B)$ (equivalently $\mathrm{supp}(B)\subseteq\mathrm{supp}(A)$)
\footnote{For any Hermitian $A$ acting on $\mathcal{H}$ its kernel (support) is a subspace of $\mathcal{H}$ spanned by the eigenvectors of $A$ corresponding to its zero (nonzero) eigenvalues. Thus for any Hermitian $A$ it holds that $\mathrm{ker}(A)=(\mathrm{supp}(A))^{\perp}$.}

To see it explicitly let $\ket{\psi}\in\mathrm{ker}(A)$ and then due to the assumption we have
\begin{equation}
0\leq \bra{\psi}B\ket{\psi}\leq \bra{\psi}A\ket{\psi}=0
\end{equation}
meaning that if $\ket{\psi}\in\mathrm{ker}(A)$, then immediately $\ket{\psi}\in\mathrm{ker}(B)$. Thus, $B$ acts on $\mathrm{ker}(A)$ 
as a zero operator and therefore in what follows we can restrict our considerations to the support 
of $A$. This in turn means that we can always assume $A$ to be invertible. Assuming then that $A$ 
is invertible we can utilize the Douglas lemma saying that if $A\geq B\geq 0$ and $A>0$ then there 
exists such $R$ that $\|R\|\leq 1$ (by $\|\cdot\|$ we denote the operator norm) and
\begin{equation}\label{relation1}
\sqrt{B}=\sqrt{A}R.
\end{equation}

Now the proof is straightforward. Let us assume that $A$ is diagonal in the standard basis 
and let $U$ be a unitary operator diagonalizing $B$ in the standard basis in $\mathbb{C}^{d}$, i.e. $B=UD_{B}U^{\dagger}$, 
where $D_{B}$ is a diagonal matrix containing nonzero eigenvalues of $B$. Then from equation (\ref{relation1}) 
we infer that
\begin{equation}\label{relation2}
\sqrt{D}_{B}=U^{\dagger}\sqrt{A}\widetilde{R},
\end{equation}
where by $\widetilde{R}$ we denoted $R U$. Note that $\|\widetilde{R}\|\leq 1$ as $\|U\|\leq 1$. From equation (\ref{relation2}) we can easily 
find that
\begin{equation}
D_{B}=\widetilde{R}^{\dagger}A\widetilde{R}
\end{equation}
which in turn allows us to write
\begin{eqnarray}
\lambda_{i}(B)&=&\bra{i}D_{B}\ket{i}\nonumber\\
&=&\bra{i}\widetilde{R}^{\dagger}A\widetilde{R}\ket{i}\nonumber\\
&=&
\sum_{j}\lambda_{j}(A)\big|\bra{i}\widetilde{R}^{\dagger}\ket{j}\big|^{2}.
\end{eqnarray}

Denoting by $S_{ij}=\big|\bra{i}\widetilde{R}^{\dagger}\ket{j}\big|^{2}\geq 0$ the elements of matrix $S$, we can rewrite the above as 
the following matrix equation: $\lambda(B)=S\lambda(A)$. Now it suffices to show that elements of $S$ obey 
the conditions for double substochasticity. For this purpose, let us notice that since $\|\widetilde{R}\|\leq 1$ then
also $\|\widetilde{R}^{\dagger}\|\leq 1$ implying that $\bra{\psi}\widetilde{R}^{\dagger}\widetilde{R}\ket{\psi}\leq 1$ and $\bra{\psi}\widetilde{R}\widetilde{R}^{\dagger}\ket{\psi}\leq 1$ hold for any $\ket{\psi}\in\mathbb{C}^{d}$. This 
allows us to write
\begin{eqnarray}
\sum_{i}S_{ij}&=&\sum_{i}\big|\bra{i}\widetilde{R}^{\dagger}\ket{j}\big|^{2}\nonumber\\
&=&\sum_{i}\bra{j}\widetilde{R}\proj{i}\widetilde{R}^{\dagger}\ket{j}\nonumber\\
&=&\bra{j}\widetilde{R}\widetilde{R}^{\dagger}\ket{j}\nonumber\\
&\leq& 1.
\end{eqnarray}

In the same way we prove that $\sum_{j}S_{ij}\leq 1$ concluding the proof $\fullsquare$.

The above fact may be also proven in a more straightforward way. Namely, it suffices 
to utilize the fact that if $A\geq B$, then $\lambda^{\downarrow}(A)\geq \lambda^{\downarrow}(B)$ (see e.g. \cite{Marshall}) from which the weak 
majorization follows. We decided, however, to keep an alternative proof as it relates directly the operator inequality $A\geq B$ and the submajorization relation. On the other hand this is yet another proof.

As a corollary of the above analysis we have the fact that if for some bipartite state $\varrho$ 
(possibly, but not necessarily separable) it holds that $\Theta_{1}(\varrho)\geq \Theta_{2}(\varrho)$ (recall that $\Theta_i=I\otimes \Lambda_i$), 
then
\begin{equation}\label{relation3}
\lambda(\Theta_{1}(\varrho))\;{}_{w}\!\!\succ\lambda(\Theta_{2}(\varrho)),
\end{equation}
i.e. eigenvalues of $\Theta_{1}(\varrho)$ submajorize eigenvalues of $\Theta_{2}(\varrho)$. We can look at this result as a form 
of generalization of majorization relation to other positive maps than the reduction one. 
Obviously, in the case of $\Lambda_{r}$ the above relation gives a rather weak criterion for separability 
as it reads $\lambda(\varrho_{A}\ot\mathbbm{1}_{d})\;{}_{w}\!\!\succ\lambda(\varrho)$. However, for states with maximally mixed subsystem  $A$, i.e.
$\varrho_{A}=(1/d)\mathbbm{1}_{d}$, the above criterion is equivalent to the Nielsen--Kempe majorization criterion
$\lambda(\varrho_{A})\succ\lambda(\varrho )$ (the same holds for $\varrho_{B}$). This easily follows from the fact that in this case both 
criteria can be violated only by the largest eigenvalue of $\varrho$. On the other hand, even if in 
general weaker for the reduction map, the above criterion allows to employ other positive 
maps, even those detecting bound entanglement. What is then important, as we will see later, is
that the criterion (\ref{relation3}) detects PPT entangled states when derived from indecomposable positive 
maps. Moreover, it induces some scalar separability criteria detecting PPT entangled states and
allowing for experimental realization.

Let us now apply the above submajorization relations to derive the aforementioned entropic 
inequalities. For this purpose, we can utilize the following known fact (see e.g. \cite{Marshall}).\\

\noindent {\it {\bf Fact 2.} Let $x$ and $y$ be two vectors from $\mathbb{R}^{d}$. If $x\;{}_{w}\!\!\succ y$ then for any Schur--convex increasing 
function $\phi:\mathbb{R}\to\mathbb{R}$, the following inequality:
\begin{equation}\label{Schur_order}
\phi(x)\geq \phi(y)
\end{equation}
is satisfied. For Schur--concave decreasing functions the sign of inequality should be changed to "$\leq $".}\\

\noindent Note, however, that using the property that if $A\geq B$ then $\lambda_i(A)\geq\lambda_i(B)$, one can derive a scalar inequality for any monotonic function, without the notion of Schur convexity/concavity.
However, in what follows we will connect the entropic criteria to week majorization as in the case of standard entropic inequalities.

The examples of already known functions that are Schur--convex/concave and at the same 
time increasing/decreasing are summarized in Table \ref{schur} together with their operator counterparts. 
If arguments of a function represent a probability distribution, then some of the presented 
functions lead to known entropies. This is also remarked in the caption to table \ref{schur}.
\begin{center}
\begin{table}[h]
  \begin{tabular}[t]  {p{3.5cm}| c |p{5.7cm}}
  \hline\hline
  \qquad function & operator counterpart & \qquad remarks \\
  \hline\hline
  $\displaystyle f_{\alpha}(x)=\sum_{i}x_{i}^{\alpha}$ \B &
  $\Tr X^{\alpha}$ &
\begin{minipage}[l]{\textwidth}
  \ \\
  $\alpha\in [0,1)$ -- Schur--concave, increasing\\
  $\alpha\geq 1$ --  Schur--convex, increasing;\\
  \end{minipage}\B \\
  \hline
  $\displaystyle h_{\alpha}^{R}(x)=\frac{\ln [f_{\alpha}(x)]}{1-\alpha}$ \TT \B &
  $\displaystyle \frac{\ln \Tr X^{\alpha}}{1-\alpha}$ &
\begin{minipage}[c]{\textwidth}
  \ \\
  $\alpha\in[0,1)$ -- Schur--concave, increasing;\\
  $\alpha\geq 1$ -- Schur--concave, decreasing;
  \end{minipage} \\
  \hline
  $\displaystyle h_{\alpha}^{T}(x)=\frac{f_{\alpha}(x)-1}{1-\alpha}$ \TT \B &
  $\displaystyle \frac{\Tr X^{\alpha}-1}{1-\alpha}$ &
\begin{minipage}[c]{\textwidth}
  \ \\
  $\alpha\in[0,1)$ -- Schur--concave, increasing; \\
  $\alpha\geq 1$ -- Schur--concave, decreasing;
  \end{minipage} \\
  \hline
  $\displaystyle h_{\alpha}^{A}(x)\!=\!\frac{\left(f_{1/\alpha}(x)\right)^{\alpha}\!\!-1}{\alpha-1}$ \T &
  $\displaystyle \frac{\left(\Tr X^{1/\alpha}\right)^{\alpha}-1}{\alpha-1}$ &
\begin{minipage}[c]{\textwidth}
  $\alpha\in[0,1)$ -- Schur--concave, decreasing;\\
  $\alpha\geq 1$ -- Schur--concave, increasing;
  \end{minipage}\\
\hline\hline
\end{tabular}
  \caption{Examples of Schur--convex (Schur--concave) and increasing (decreasing)
functions defined on $\mathbb{R}_{+}^n$ and their counterparts
defined on the set of positive matrices. Functions are defined for
$\alpha\geq 0$ and their properties depending on the range of
$\alpha$ are given. The functions $h_{\alpha}^{R}$,
$h_{\alpha}^{T}$, and $h_{\alpha}^{A}$ correspond to the Renyi
\cite{Renyi}, Tsallis \cite{Tsallis}, and Arimoto \cite{Arimoto}
entropies, respectively. If the argument of a function is a
probability distribution (or a normalized, positive matrix) the
functions give exactly the mentioned entropies or their quantum
counterparts (see e.g. \cite{Wehrl}).}\label{schur}
\end{table}
\end{center}

For instance, using the function $f_{\alpha}$ with $\alpha\geq 1$ (see Table \ref{schur}) we can easily obtain the
inequality
\begin{equation}\label{nierownosc}
\Tr\left[\Theta_1(\varrho)\right]^{\alpha} \geq \Tr\left[\Theta_2(\varrho)\right]^{\alpha}.
\end{equation}
Note that the inequality can be written also for $\alpha\in[0,1)$, as in such case the operator function
$\psi(A)=A^{\alpha}$ is monotonically increasing (see e.g. \cite{Bhatia}). Now, either from inequality (\ref{nierownosc}) or 
directly using the entropic functions, we can obtain e.g.
\begin{equation}\label{nierownosc2}
S_{\alpha}^{R(T)}(\Theta_{1}(\varrho))\leq
S_{\alpha}^{R(T)}(\Theta_{2}(\varrho)).
\end{equation}
Let us also notice that in the case of the Renyi entropy, taking the limit of $\alpha\to\infty$ on both sides 
of equation (\ref{nierownosc2}) leads to
\begin{equation}\label{normInequality}
\|\Theta_{1}(\varrho )\|\geq \|\Theta_{2}(\varrho )\|.
\end{equation}
(Note that in the case of positive matrix the operator norm is just its largest eigenvalue.) We 
also used the fact that $\lim_{\alpha\to\infty}S_{\alpha}^{R}(\varrho)=-\ln\|\varrho\|$. This inequality is also a straightforward conclusion following from the assumption that $\Theta_{1}(\varrho )\geq \Theta_{2}(\varrho )$ and the fact that if $A\geq B$ then
$\|A\|\geq \|B\|$.

The above results will be discussed in a more details in Section \ref{Special}. Now we present yet another method allowing to derive different inequalities.

\subsection{Derivation of scalar separability criteria based on functional inequalities}

As a continuation of the work \cite{RAJS} we provide below a slightly different method that 
allows us to derive some entropic inequalities. The new inequalities are more general than the 
ones studied in \cite{RAJS}, as one does not have to impose on the density matrices any restrictions 
in the form of commutation relations. Moreover, they serve better as a separability criterion. It 
should be also stated that in some particular case, they resemble the just derived inequalities, 
but are stronger. We will show the relation between them in the section concerning special cases of both inequalities.

Let us start by proving the following.\\

\noindent {\it {\bf Theorem 2.} Let $\Lambda$ be some positive map
such that $[I\ot\Lambda](\varrho )\geq 0$ for some $\varrho $.
Then the following inequalities hold.

\begin{description}
   \item[(i)] For $\alpha,\beta\geq 0$, we have
\begin{equation}\label{TheoremIneq1}
\Tr\left\{[\Theta_{1}(\varrho )]^{\alpha}[\Theta_{2}(\varrho
)]^{\beta}\right\}\geq \Tr[\Theta_{2}(\varrho )]^{\alpha+\beta}.
\end{equation}
   \item[(ii)] If $\beta\geq 0$ and $\alpha\in(0,1]$, then
\begin{equation}\label{TheoremIneq2}
\Tr\left\{[\Theta_{1}(\varrho )]^{-\alpha}[\Theta_{2}(\varrho
)]^{\beta}\right\}\leq \Tr[\Theta_{2}(\varrho )]^{-\alpha+\beta}.
\end{equation}
  \item[(iii)] In the particular case of $\alpha\to\infty$, the
inequalities (\ref{TheoremIneq1}) lead to the inequality
\begin{equation}\label{Qmax}
q_{\max}\geq \|\Theta_{2}(\varrho )\|,
\end{equation}
where $q_{\max}$ is maximal of the eigenvalues $\{\lambda_i\}$ of
$\Theta_{1}(\varrho )$ corresponding to nonzero mean values
$\bra{\psi^{(1)}_{i}}\Theta_{2}(\varrho )\ket{\psi^{(1)}_{i}}$,
where $\{\ket{\psi^{(1)}_{i}}\}$ are the eigenvectors of
$\Theta_{1}(\varrho )$.
\end{description}
}

\noindent {\it Proof.} We can easily follow the method presented in \cite{VollbrechtWolf}. Firstly let us note that by the
assumption $\Theta_{1}(\varrho )\geq \Theta_{2}(\varrho )$. Therefore, we have
\begin{equation}\label{nierownoscOp}
\Theta_{1}(\varrho )+\epsilon\mathbbm{1}_{d}\geq
\Theta_{2}(\varrho )+\epsilon\mathbbm{1}_{d}
\end{equation}
for any $\epsilon>0$. Then we can write
\begin{eqnarray}
&&\Tr\left\{\left[\Theta_{1}(\varrho
)+\epsilon\mathbbm{1}_{d}\right]^{\alpha}
\left[\Theta_{2}(\varrho )+\epsilon\mathbbm{1}_{d}\right]^{\beta}\right\}\\
&&\hspace{2cm}=\Tr\left\{\mathrm{e}^{\alpha\ln[\Theta_{1}(\varrho
)+\epsilon\mathbbm{1}_{d}]}
\,\mathrm{e}^{\alpha\ln[\Theta_{2}(\varrho
)+\epsilon\mathbbm{1}_{d}]}\right\}.\nonumber
\end{eqnarray}
Then, due to the Golden--Thompson inequality (cf \cite{Bhatia}) saying that $\Tr\,\mathrm{e}^{A}\mathrm{e}^{B}\geq
\Tr\,\mathrm{e}^{A+B}$, we can write
\begin{eqnarray}
&&\Tr\left\{\left[\Theta_{1}(\varrho
)+\epsilon\mathbbm{1}_{d}\right]^{\alpha}
\left[\Theta_{2}(\varrho )+\epsilon\mathbbm{1}_{d}\right]^{\beta}\right\}\\
&&\hspace{2cm}\geq \Tr\,\mathrm{e}^{\alpha\ln[\Theta_{1}(\varrho
)+\epsilon\mathbbm{1}_{d}]+\beta\ln[\Theta_{2}(\varrho
)+\epsilon\mathbbm{1}_{d}]}.\nonumber
\end{eqnarray}
Now, we can utilize the assumption, the fact that $\Tr\,\mathrm{e}^{A}\geq \Tr\,\mathrm{e}^{B}$ for $A\geq B$
\cite{VollbrechtWolf}, and the monotonicity of logarithm ($A\geq B>0$ then $\log A\geq \log B$) \cite{Lowner},obtaining
\begin{eqnarray}
&&\Tr\left\{\left[\Theta_{1}(\varrho
)+\epsilon\mathbbm{1}_{d}\right]^{\alpha}
\left[\Theta_{2}(\varrho )+\epsilon\mathbbm{1}_{d}\right]^{\beta}\right\}\nonumber\\
&&\hspace{2cm}\geq \Tr\,\mathrm{e}^{(\alpha+\beta)\log[\Theta_{2}(\varrho )+\epsilon\mathbbm{1}_{d}]}\nonumber\\
&&\hspace{2cm}=\Tr[\Theta_{2}(\varrho
)+\epsilon\mathbbm{1}_{d}]^{\alpha+\beta}.
\end{eqnarray}
To finish the proof of the first inequality it suffices to take limit $\epsilon\to 0$ on both sides of the above.

To proceed with the second inequality we follow similar technique as in \cite{Lindblad} and 
utilize the fact that the operator function $g(A)=A^{r}$ is monotonically decreasing for $r\in[-1,0)$. 
The latter after application to the inequality (\ref{nierownoscOp}) allows us to write
\begin{equation}
(\Theta_{1}(\varrho)+\epsilon\mathbbm{1}_{d})^{-\alpha}\leq
(\Theta_{2}(\varrho)+\epsilon\mathbbm{1}_{d})^{-\alpha}
\end{equation}
for any $\epsilon>0$. Since it holds that if $A\geq B$ then $XAX^{\dagger}\geq XBX^{\dagger}$ for any matrix $X$ (cf \cite{Bhatia}), we have
\begin{eqnarray}
&&[\Theta_{2}(\varrho)]^{\beta/2}P_{2}(\Theta_{1}(\varrho
)+\epsilon\mathbbm{1}_{d})^{-\alpha}P_{2}[\Theta_{2}(\varrho)]^{\beta/2}\nonumber\\
&&\hspace{2cm}\leq [\Theta_{2}(\varrho)]^{\beta/2}P_{2}(\Theta_{2}(\varrho
)+\epsilon\mathbbm{1}_{d})^{-\alpha}
P_{2}[\Theta_{2}(\varrho)]^{\beta/2},
\end{eqnarray}
where by $P_{2}$ we denoted the projector onto the support of $\Theta_{2}(\varrho)$ (it is necessary when $\alpha>\beta$). Taking the trace on both sides we obtain
\begin{eqnarray}
&&\Tr\left\{[\Theta_{2}(\varrho
)]^{\beta}(\Theta_{1}(\varrho
)+\epsilon\mathbbm{1}_{d})^{-\alpha}
\right\}\\
&&\hspace{2cm}\leq\Tr\left\{[\Theta_{2}(\varrho
)]^{\beta}P_{2}(\Theta_{2}(\varrho
)+\epsilon\mathbbm{1}_{d})^{-\alpha}P_{2}\right\}.\nonumber
\end{eqnarray}
To finish it suffices to take the limit $\epsilon\to\infty$ from both sides remembering that for $\alpha>\beta$ one 
has to take the pseudoinverse of $\Theta_{2}(\varrho)$ (thus we put $P_{2}$ here). This finishes the proof of this part.

Finally, we discuss the behaviour of the inequalities (\ref{TheoremIneq1}) in the limit of $\alpha\to\infty$. 
Let $\gamma_{i}=\bra{\psi_i^{(1)}}[\Theta_{2}(\varrho)]^{\beta}\ket{\psi_i^{(1)}}$, where $\ket{\psi_{i}}\in\mathrm{supp}[\Theta_{1}(\varrho )]$. Then the inequality (\ref{TheoremIneq1}) can be stated as
\begin{equation}
\sum_{i}\left(\lambda^{(1)}_{i}\right)^{\alpha}\gamma_{i}\geq
\sum_{i}\left(\lambda^{(2)}_{i}\right)^{\alpha+\beta},
\end{equation}
where $\lambda_{i}^{(j)}>0$ are nonzero eigenvalues of $\Theta_{j}(\varrho )$ $(j=1,2)$. Now, we can take the logarithm from both sides and divide the whole inequality by $1-\alpha$ $(\alpha>1)$, thereby getting
\begin{equation}\label{proof3}
\frac{1}{1-\alpha}\ln\left[\sum_{i}\left(\lambda^{(1)}_{i}\right)^{\alpha}\gamma_{i}\right]\leq
\frac{1}{1-\alpha}\ln\left[\sum_{i}\left(\lambda^{(2)}_{i}\right)^{\alpha+\beta}\right].
\end{equation}
Eventually, we take the limit $\alpha\to \infty$ and obtain the required relation. This finishes the proof of the third part and the whole theorem.
$\fullsquare$

A simple example given below shows that taking $q_{\mathrm{max}}$ instead of $\|\Theta_{1}(\varrho )\|$ on the left--hand 
side of (\ref{Qmax}) is indeed necessary. As an illustrative example let us consider a pure separable state $\ket{\psi}=\ket{\psi_A}\ket{\psi_B}\in\mathbb{C}^{d}\ot\mathbb{C}^{d}$ and a transposition map $T$. For any $d\times d$ matrix $X$ one can write $T(X)$ as
\begin{eqnarray}\label{TranspDec}
T(X)=\underbrace{\frac{1}{2}\left[\Tr(X)\mathbbm{1}_{d}+T(X)\right]}&-&
\underbrace{\frac{1}{2}\left[\Tr(X)\mathbbm{1}_{d}-T(X)\right]}\nonumber\\
\hspace{2.5cm}T_{1}(X) & & \hspace{1.2cm} T_{2}(X)\nonumber\\
\end{eqnarray}
where both $T_1$ and $T_2$ are completely positive (one recognizes that $T_2$ after normalization corresponds to the
Werner--Holevo channel $\Phi_{WH}(X)=[1/(d-1)][\Tr(X)\mathbbm{1}_{d}-T(X)]$ \cite{WernerHolevo}). 
Acting with $T_{1}$ and $T_{2}$ on the second subsystem of $\ket{\psi}$, we obtain
\begin{equation}
[I\ot T_{1(2)}](\proj{\psi})=\frac{1}{2}\left(\proj{\psi_A}\otimes
\mathbbm{1}_{d}\pm\proj{\psi_A}\otimes\proj{\psi_B^{*}}\right),
\end{equation}

The largest eigenvalue of $[I\otimes T_1](\proj{\psi})$ is $\lambda_{\mathrm{max}}=1$ (not degenerated), which 
corresponds to the eigenvector $\ket{\psi_{1}^{(1)}}=\ket{\psi_{A}}\ket{\psi_{B}^{*}}$. However, the average $\gamma_1=\bra{\psi_{1}^{(1)}}[I\otimes T_2](\proj{\psi})\ket{\psi_{1}^{(1)}}$ 
equals zero and the term $\lambda_{\mathrm{\max}}^{\alpha}\gamma_{1}$ does not contribute to the left hand side 
of (\ref{proof3}). Therefore, $\lambda_{\mathrm{max}}=\|\Theta_{1}(\varrho)\|$ cannot be the limit of it.

So far we have derived scalar inequalities constituting necessary criteria for separability. 
In what follows we show how to obtain from them the formulas involving entropies and discuss 
some of the derived inequalities in the context of state merging protocol and approximation 
of a mean value of the `tailor--made' linear entanglement witness.

\subsection{Special cases}
\label{Special}

First, let us check what is the relation between the inequalities derived from the reduction map 
$\Lambda_{R}$ with both methods (i.e. Eq. (\ref{nierownosc2}) and (\ref{TheoremIneq1}), (\ref{TheoremIneq2})), and standard entropic inequalities. In 
this case $\Lambda_{1}(X)=\Tr(X)\mathbbm{1}_{d}$ and $\Lambda_{2}=I$. Thus, the inequality (\ref{nierownosc}) gives
\begin{equation}\label{red}
d\Tr\varrho_{A}^{\alpha}\geq \Tr\varrho ^{\alpha} \qquad
(\alpha\geq 0).
\end{equation}
Comparing to the standard entropic inequalities (\ref{entrWithout1}), we see that the first method leads in 
this case to weaker inequalities for $\alpha\geq 1$.

The second method applied in this paper brought us to equations (\ref{TheoremIneq1}) and (\ref{TheoremIneq2}). For the
reduction map we get the following:
\begin{equation}
\Tr\left[(\varrho_{A}\ot \mathbbm{1}_{d})^{\alpha}\varrho^{\beta}\right]\geq \Tr\varrho ^{\alpha+\beta}\qquad (\alpha,\beta\geq 0)
\end{equation}
and
\begin{equation}
\Tr\left[(\varrho_{A}\ot \mathbbm{1}_{d})^{\alpha}\varrho^{\beta}\right]\leq \Tr\varrho_{AB}^{\alpha+\beta}\qquad (-1\leq \alpha<0,\beta\geq 0).
\end{equation}
Now putting $\beta=1$ we reproduce the standard entropic inequalities (\ref{entrWithout1}) and (\ref{entrWithout2}), respectively. In conclusion, in the case of the reduction map, the first method does not reproduce the standard
entropic inequalities as a particular example, while the second method does.

In general, it is hard to investigate the effectiveness of our inequalities versus positive 
maps as they can have many decompositions of the form (\ref{relation5}) and the choice of decomposition 
strongly affects the effectiveness of the inequalities. For instance, the Breuer--Hall map can be 
written as in Eq. (\ref{relation5}) with $\Lambda_{1}^{(1)}=\Lambda_{\mathrm{Tr}}-\tau_{U}-(1/2) I$ and $\Lambda_{2}^{(1)}=(1/2) I$ or $\Lambda_{1}^{(2)}=2\Lambda_{\mathrm{Tr}}$ and $\Lambda_{2}^{(2)}=\Lambda_{\mathrm{Tr}}+I+\tau_{U}$. One easily checks complete positivity of above maps applying 
the corollary from Choi-Jamio\l{}kowski isomorphism \cite{Jam,Choi_iso}, known as Jamio\l{}kowski criterion for complete positivity.

We will present the particular decomposition, which naturally arises in the case of the 
reduction map, and which may be applied to any positive map. Let us then utilize the following 
fact (see e.g. \cite{SperlingVogel}).\\

\noindent {\it {\bf Fact 3.} Let $\Lambda:M_{d}(\mathbb{C})\to M_{d}(\mathbb{C})$ be a positive map. Then $\Lambda$ can always be decomposed in the following way:
\begin{equation}\label{PMdecomp}
\Lambda(X)=\xi \Tr(X)\mathbbm{1}_{d}-\Lambda_{2}(X),
\end{equation}
where $\xi=d\lambda_{\max}$ with $\lambda_{\max}$ denoting maximal eigenvalue of $[I \ot\Lambda](P_{+}^{(d)})$, and $\Lambda_{2}$ is some completely positive map.}\\

\noindent {\it Proof.} This fact may be proven as Lemma 1 in \cite{SperlingVogel}. For completeness, we provide here this
reasoning translated to maps by the Choi--Jamio\l{}kowski isomorphism \cite{Jam,Choi_iso}. We can write the positive map $\Lambda$ as $\Lambda=\xi\Lambda_{\mathrm{Tr}}-(\xi\Lambda_{\mathrm{Tr}}-\Lambda)$. Then, since $\Lambda_{\mathrm{Tr}}$ is completely positive it suffices to show that $\Lambda_{2}=(\xi\Lambda_{\mathrm{Tr}}-\Lambda)$ is completely positive. Using the Jamio\l{}kowski criterion for complete positivity, one gets
\begin{eqnarray}
d[I\ot \Lambda_{2}](P_{+}^{(d)})&=&\xi
\mathbbm{1}_{d}\ot\mathbbm{1}_{d}-[I\ot
\Lambda](P_{+}^{(d)})\nonumber\\
&=&\sum_{i}(\xi-d\lambda_{i})\proj{\psi_{i}}\nonumber\\
&\geq &0.
\end{eqnarray}
The last inequality is a consequence of the fact that $\xi\equiv d\max_{i}\lambda_{i}\geq d\lambda_{i}$ for any $i$. Here by $\{\lambda_{i},\ket{\psi_{i}}\}$ we denoted the spectral decomposition of $[I\ot\Lambda](P_{+}^{(d)})$ including the zero eigenvalues. $\fullsquare$

This means that we can always decompose a positive map onto the difference of $\Lambda_{\mathrm{Tr}}$ 
multiplied by some factor and some other completely positive map. In this way, we can fix 
one of the maps appearing in the decomposition (\ref{relation5}). We go even further in simplifying our 
considerations. Namely, we restrict our attention to such maps that $\Lambda_{2}$ can be normalized, i.e. $\Tr[\Lambda_{2}(X)]=\eta_{d}\Tr(X)$ for all $X\geq 0$. Therefore, since $\Lambda_{2}$ is completely positive, dividing it by $\eta_{d}$ leads to some quantum channel $\Phi=(1/\eta_{d})\Lambda_{2}$. In other words, we consider only the maps that can be written as 
\begin{equation}\label{decompositionEta}
\Lambda(X)=\xi \Tr (X)\mathbbm{1}_{d}-\eta_{d}\Phi(X).
\end{equation}
Many maps known from the literature admit the above form, for instance, the generalization of transposition map $\tau^{U}(X)$ ($U$ denotes some unitary matrix); reduction map $\Lambda_{R}$ \cite{reduction,CerfAdamiGingrich} and the map introduced by Breuer and Hall $\Lambda_{BH}^{U}$ \cite{Breuer,Hall}; and finally, the Choi map \cite{Choi} and some of its generalizations studied in \cite{uogolnieniaChoi,Ha1}, namely,
\begin{equation}
\tau_{d,k}(X)=\epsilon(X)+\sum_{i=1}^{k}\epsilon(S^{i}XS^{i\dagger})-X,
\end{equation}
where $\epsilon$ denotes the completely positive map defined as
\begin{equation}
\epsilon(X)=\sum_{j=0}^{d-1}\bra{j}X\ket{j}\proj{j}
\end{equation}
and $S\ket{i}=\ket{i+1}$ (mod $d$). For $k=1,\ldots,d-2$, the maps $\tau_{d,k}$ were shown to be 
indecomposable \cite{Ha1}. For $k=0$, one has a completely positive map, $k=d-1$ reproduces the
reduction map $\Lambda_{R}$, and $\tau_{3,1}$ is the Choi map \cite{Choi}. Decompositions of the form (\ref{decompositionEta}) for the above positive maps are summarized in Table \ref{maps}.

With such decomposition we obtain inequalities involving entropies which resemble the 
standard once. To show this let us concentrate on the inequalities (\ref{nierownosc2}) and fix the entropies to be 
the Renyi entropy. If for some $\varrho_{AB}$ (possibly, but not necessarily separable) acting on $\mathbb{C}^{d}\ot\mathbb{C}^{d}$ it holds that $[I\ot\Lambda](\varrho )\geq 0$, then by virtue of the decomposition (\ref{decompositionEta}) we get the following relation:
\begin{equation}
S_{\alpha}(\xi\varrho_{A}\ot\mathbbm{1}_{d})\leq
S_{\alpha}(\eta_{d}[I\ot\Phi](\varrho )),
\end{equation}
which due to the following facts:
$S_{\alpha}(\xi\varrho_{A}\ot\mathbbm{1}_{d})=[1/(1-\alpha)]\ln\xi^{\alpha}d+S_{\alpha}(\varrho_{A})$
and $S_{\alpha}(\eta_d [I\otimes\Phi](\varrho
))=[1/(1-\alpha)]\ln\eta_d^{\alpha}+S_{\alpha}([I\otimes\Phi](\varrho
))$ gives after a little algebra
\begin{equation}\label{derivedIn}
S_{\alpha}^{R}([I\ot\Phi](\varrho
))-S_{\alpha}^{R}(\varrho_{A})\geq\ln\frac{\eta_d}{\xi}- \frac{\ln
(d\xi /\eta_d)}{\alpha-1},
\end{equation}
for $\alpha\geq 0$. What we got here has the form of the standard entropic inequality, but with 
subsystem $B$ passed through the quantum channel $\Phi$ instead of $\varrho$ on the left--hand side. Another 
difference is that at least for the maps presented in Table \ref{maps} (except the reduction one) it holds that $\eta_{d}> \xi$ and $\xi d\geq \eta_{d}$. Thus, the term appearing on the right hand side is positive and 
the second one decreases with $\alpha\to\infty$, which makes the right--hand side positive for $\alpha$ large enough and the inequality stronger than the standard entropic one for $[I\otimes\Phi](\varrho)$.
In case of the reduction map, we can see again that the present inequality is weaker than standard entropic inequality since the term appearing on right--hand side is negative. In the limit $\alpha\to\infty$ we get, similarly to equation (\ref{normInequality}),
that
\begin{equation}\label{krytNormy}
\|\varrho_{A}\|\geq \frac{\eta_{d}}{\xi}\left\|[I\ot\Phi](\varrho
)\right\|.
\end{equation}
\begin{center}
\begin{table}[h]
  \centering
  \begin{tabular}
  {p{4cm}| c |p{3.6cm}|c}
    \hline\hline
  \qquad map \BB & $\xi$ & $\Lambda_2(X)$ & $\eta_d$ \\
  \hline\hline
  Transposition \TTT $\displaystyle \tau^U(X)$ &\ $1$ \ & $ \Tr(X)\mathbbm{1}-\tau^U(X)$ & $d-1$\\[1.5ex]
  Breuer--Hall $\displaystyle \Lambda_{BH}(X)$ &\ $2$ \ & $ \Tr(X)\mathbbm{1}+X+\tau^V(X)$ & $d+2$\\[1.5ex]
  Reduction $\displaystyle \Lambda_R(X)$ &\ $1$ \ & $X$ & $1$ \\[1.5ex]
  Generalized Choi $\displaystyle \tau_{d,k}(X)$, $k=1,\ldots,d-2$ &\ $d-k$ \ & $(d-k)\Tr(X)-\tau_{d,k}(X)$ & $d(d-k)-d+1$ \\[1.5ex]
    \hline\hline
\end{tabular}
  \caption{Summary of the most common positive maps and their decompositions with $\Lambda_1$ proportional to $\Tr(X)\mathbbm{1}$.}\label{maps}
\end{table}
\end{center}

Let us now move to the inequalities proven in Theorem 2. As one can easily find, putting 
$\beta=1$ and taking the decomposition (\ref{decompositionEta}), we obtain from both inequalities, (\ref{TheoremIneq1}) and (\ref{TheoremIneq2}), the following one:
\begin{equation}\label{inequalitiesStr3}
S_{\alpha}^{R}([I\ot\Phi](\varrho))-S_{\alpha}^{R}(\varrho_{A})\geq\ln\frac{\eta_{d}}{\xi}\qquad (\alpha\geq 0).
\end{equation}
Again, as in the case of equation (\ref{derivedIn}) the term on the right--hand side is positive for the 
positive maps presented in Table \ref{maps} (except the reduction one for which the term equals zero). 
Moreover, contrary to the previous example, the right--hand side does not depend on $\alpha$. One can 
immediately establish a relation between (\ref{derivedIn}) and the present case, i.e. for such maps that the 
term $\ln (d\xi/\eta_d)$ is positive and for $\alpha > 1$, the inequality (\ref{inequalitiesStr3}) constitutes a stronger separability 
criterion than (\ref{derivedIn}). However, in the limit $\alpha\to \infty$ they are equivalent.

The fact that the right--hand side of (\ref{inequalitiesStr3}) is independent of $\alpha$ can lead to interesting
conclusion. Namely, applying the limit $\lim_{\alpha\to 1}$ and utilizing the fact that $\lim_{\alpha\to 1}S_{\alpha}^{R}(\varrho)=S(\varrho)$, we get the following inequality for the von Neumann entropy:
\begin{equation}\label{niervonN}
S([I\ot\Phi](\varrho ))-S(\varrho_{A})\geq\ln\frac{\eta_{d}}{\xi}.
\end{equation}

One knows that the conditional von Neumann entropy $S(B|A;\varrho)=S(\varrho )-S(\varrho_{A})$ is a 
minimal amount of quantum communication necessary to merge a quantum state \cite{MHJOAW}. For 
separable states, it is always larger or equal zero. However, for some entangled states, 
violating the standard entropic inequality, the cost can be negative. This means 
that one can, actually, extract some entanglement in the protocol of state merging and use it for 
future quantum communication. Now application of some map to a separable state increases the 
lower bound on the cost of state merging, i.e. merging a still separable state $[I\ot \Phi](\varrho) $ costs 
not less than $\ln(\eta_{d}/\xi)$. In this way we provide a lower bound for the cost of merging a separable 
state after local action of a quantum channel.

On the other hand for states that are entangled and detected by the inequality (\ref{niervonN}) we know 
that the cost of merging a state after partial action of the quantum channel $\Phi$ would be smaller 
than the bound $\ln(\eta_{d}/\xi)$.

Moreover, there exist states for which the conditional entropy is negative, however the 
inequality (\ref{niervonN}) is not violated. This means that the channel destroys quantum correlations in 
such way that extracting entanglement in the protocol of state merging becomes impossible.

\noindent{\it Example.} As an example of states possessing such feature let us consider the following 
rotationally invariant $4\ot 4$ density matrices:
$\sigma=p P_0+(1-p) P_1$.
It is a mixture of two 
operators projecting on the eigenspaces of the square of total angular momentum corresponding 
to $J=0$ and $J=1$, and normalized to have trace one. We denote them, respectively, by $P_0$ 
and $P_1$. 
The state is entangled for any value of parameter $p$ and its conditional von Neumann 
entropy is negative for all $p$ except for $p=1/4$. After one of the subsystems is passed through 
the Werner--Holevo channel, the conditional entropy of the state becomes positive, however for 
states with $p\approx 0.535$ and higher, it still violates inequality (\ref{niervonN}), so the conditional entropy is 
not larger than $\ln 3$.

Let us finally discuss the above criteria for the special class of states $\varrho $, i.e. those that have 
at least one maximally mixed subsystem (e.g. $\varrho_{A}=(1/d)\mathbbm{1}_{d}$). Note that for states that fulfill the
weaker condition, having at least one full--rank subsystem (let us assume subsystem $A$), we can 
do the transformation called local filtering. Acting with $(\varrho_{A})^{-1/2}$ on $A$, we obtain
\begin{equation}
\varrho' =\sqrt{\varrho_{A}^{-1}}\ot\mathbbm{1}_{d}\varrho \sqrt{\varrho_{A}^{-1}}\ot\mathbbm{1}_{d}
\end{equation}
The output state $\varrho'$ has a maximally mixed subsystem and the same separability properties as an 
original state ($\varrho$ is separable iff $\varrho'$ is).

For separable states with at least one maximally mixed subsystem, we can show the 
relation between two cases of inequality (\ref{TheoremIneq1}), namely one with $\alpha\geq 1, \beta=1$ and the other with 
$\alpha= 1, \beta\geq 1$, both derived from decomposition (\ref{decompositionEta}). First note that in such a case, we write the 
inequalities as follows:
\begin{equation}\label{aeq1}
\frac{\xi}{d}\Tr[\Theta_2(\varrho)]^{\beta-1} \geq \Tr[\Theta_2(\varrho)]^{\beta}\qquad (\alpha=1)
\end{equation}
and
\begin{equation}\label{beq1}
\left(\frac{\xi}{d}\right)^{\alpha}\Tr[\Theta_2(\varrho)] \geq \Tr[\Theta_2(\varrho)]^{\alpha+1}\qquad (\beta=1).
\end{equation}
On the left hand side of inequality (\ref{aeq1}) appears a term $\Tr[\Theta_{2}(\varrho)]^{\beta-1}$, which due to the same
inequality is bounded from above as
\begin{equation}
\Tr[\Theta_2(\varrho)]^{\beta-1} \leq \frac{\xi}{d}\Tr[\Theta_2(\varrho)]^{\beta-2}.
\end{equation}
In this way from equation (\ref{aeq1}) we obtain the sequence of inequalities:
\begin{eqnarray}
\Tr[\Theta_2(\varrho)]^{\beta} &\leq&
\frac{\xi}{d}\Tr[\Theta_2(\varrho)]^{\beta-1}\nonumber\\
&\leq& \cdots \leq
\nonumber\\
&\leq & \left(\frac{\xi}{d}\right)^{\beta-1}
\Tr[\Theta_2(\varrho)].
\end{eqnarray}

Taking the first and the last term of the above sequence and changing $\beta$ to $\alpha+1$, we obtain 
inequality (\ref{beq1}). In this way, we have shown that inequality (\ref{beq1}) is implied by (\ref{aeq1}) nad therefore 
for states with maximally mixed subsystem (\ref{beq1}) is always weaker. This leads to conclusion
that for these states the standard entropic inequality is always weaker than inequality (\ref{aeq1}) 
derived from reduction map. An example showing this dependence is presented in the next 
section.

Let us now state the equivalence between various criteria for states with maximally mixed 
subsystem. As already mentioned, $\varrho_A\ot\mathbbm{1}_d=(1/d) \mathbbm{1}_{d^2}$; thus the decomposition (\ref{PMdecomp}) leads in this 
case to the following:
\begin{equation}\label{PMdecMM}
(I\ot\Lambda)(\varrho)=\frac{\xi}{d}\mathbbm{1}_{d^2}-\Theta_{2}(\varrho).
\end{equation}
Note, that due to the above relation both maps, $\Theta_2(\varrho)$ and $(I\ot\Lambda)(\varrho)$, have the same
eigenvectors. Moreover, the projector corresponding to the maximal eigenvalue of $\Theta_{2}(\varrho)$ 
is the same as for the minimal eigenvalue of $(I\otimes\Lambda)(\varrho)$. Thus, in this case, the positive 
map separability criterion is equivalent to the weak majorization criterion derived from 
decomposition (\ref{PMdecMM}). This, in turn, is equivalent to the inequality involving only maximal 
eigenvalues. Precisely speaking we have the following:
\begin{eqnarray}
\hspace{0.9cm}(I\ot \Lambda)(\varrho)\geq 0&&\nonumber\\
\hspace{1.9cm}\Updownarrow&&\nonumber\\
\displaystyle\left(\frac{\xi}{d},\ldots,\frac{\xi}{d}\right){}_w\!\!\succ\lambda(\Theta_{2}(\varrho))&&\nonumber\\
\hspace{1.9cm}\Updownarrow&&\nonumber\\
\hspace{0.9cm}\displaystyle \frac{\xi}{d}\geq \|\Theta_2(\varrho)\|.&&
\end{eqnarray}

We will use the above statements to show how the new inequalities approximate a mean 
value of a linear entanglement witness in the case of states with at least one maximally mixed 
subsystem. Let us assume that an entangled state $\varrho$ is detected by the map $\Lambda$ when it acts on 
subsystem $B$, i.e. $[I\otimes\Lambda](\varrho)$ has a minimal negative eigenvalue $\lambda_{-}$. In such case we know that an entanglement witness can be constructed by acting with $I\otimes\Lambda^{\dagger}$ 
\footnote{By $\Lambda^{\dagger}$ we denoted the adjoint of $\Lambda$. Consider $M_{d}(\mathbbm{C})$ as a Hilbert space
with Hilbert--Schmidt product $(A,B)=\Tr(A^{\dagger}B)$. The adjoint map is by definition such that it satisfies the
following: $\Tr A^{\dagger} \Lambda(B)=\Tr [\Lambda^{\dagger}(A)]^{\dagger} B.$}
on the projector $P_{-}$ 
corresponding to $\lambda_{-}$, i.e.
\begin{equation}\label{witt}
\mathcal{W_{\varrho}}=[I\otimes\Lambda^{\dagger}](P_{-}).
\end{equation}
We will call such witness a "tailor--made" entanglement witness. However, assuming that one 
does not have any previous knowledge about the spectral decomposition of the analyzed state it 
seems impossible to apply such witness experimentally. Our method allows us to approximate 
the measurement of such witness on the condition that the subsystem $A$ is maximally mixed.

Let the above assumptions hold and let us consider the inequality from Theorem 2 with 
$\alpha=1$ and $\beta>1$, derived from $\Lambda$. In the case of states with maximally mixed subsystem, we 
have in view of Eq. (\ref{PMdecMM})
\begin{equation}\label{approx}
\Tr\left\{[I\otimes\Lambda](\varrho )[\Theta_2(\varrho
)]^{\beta}\right\}=\sum_{i}\lambda_{i}([I\ot\Lambda](\varrho))\,
[\lambda_i(\Theta_2(\varrho))]^{\beta},
\end{equation}
where $\lambda_i(\cdot)$ denotes the eigenvalue corresponding to the $i$th eigenvector of both $[I\otimes\Lambda](\varrho)$ and
$\Theta_2(\varrho)$. For sufficiently large $\beta$, the dominating term on the left hand side is $[\lambda_{\mathrm{max}}(\Theta_2(\varrho))]^{\beta}\lambda_{-}$ 
(recall that the maximum eigenvalue of $\Theta_2(\varrho)$ corresponds to the same projector $P_{-}$ as the
minimal negative eigenvalue of $[I\ot\Lambda](\varrho)$). Consequently, for large $\beta$ the expression (\ref{approx}) has 
the same sign as the mean value of a `tailor--made' entanglement witness (\ref{witt}), i.e. $\Tr(\mathcal{W}_{\varrho}\varrho)=\lambda_{-}$. 
As $\beta\to\infty$ the inequality detects all states detected by the map itself. In this way, if 
one could normalize $\Theta_2(\varrho)$ so that $\lambda_{\mathrm{max}}=1$ then the left hand side of inequality (\ref{approx}) would 
approximate a mean value of a `tailor--made' linear entanglement witness (\ref{witt}).

A similar approach was applied in \cite{RAJS}, however the situation considered there was slightly 
different. Namely the approximated average corresponded to Hermitian operator $[I\otimes\Lambda^{\dagger}](P_{\mathrm{max}})$ 
and $P_{\mathrm{max}}$ was a projector corresponding to the maximal eigenvalue of $\varrho$. It was not clear 
why for some states this operator can be considered an entanglement witness. In the scenario 
presented here the correspondence to entanglement witness follows explicitly from the fact that 
the projector corresponding to the maximal eigenvalue of $\Theta_{2}(\varrho)$ is the same as for the minimal 
eigenvalue of $[I\otimes\Lambda](\varrho)$.

One should note that the left--hand side of equation (\ref{approx}) can be rewritten similarly as 
in equation (\ref{aeq1}), i.e. involving the moments of $\Theta_2(\varrho)$ with power $\beta-1$ and $\beta$. However, to 
determine the full spectrum of $\Theta_2(\varrho)$ it is enough to know the first $d^2$ moments, which involves 
measuring at most $d^2$ copies at a time in a collective measurement (see \cite{manycopy} and references therein).
Therefore, one can apply the simple inequality (\ref{aeq1}) at each step of measurement of 
$d^2$ moments, and if it does not determine entanglement after one has measured $d^2-1$ and $d^2$ 
moments one should determine the spectrum of $\Theta_2(\varrho)$ and apply the inequality (\ref{krytNormy}).

All inequalities involving entropies which were given in this section can be rewritten also 
in terms of Tsallis and Arimoto entropies (see Table \ref{schur}). The formulas will be slightly different; 
however, the properties will remain the same.

\section{Effectiveness of the criteria}\label{comparison}
\begin{figure}
\centering
\includegraphics[width=0.3\textwidth]{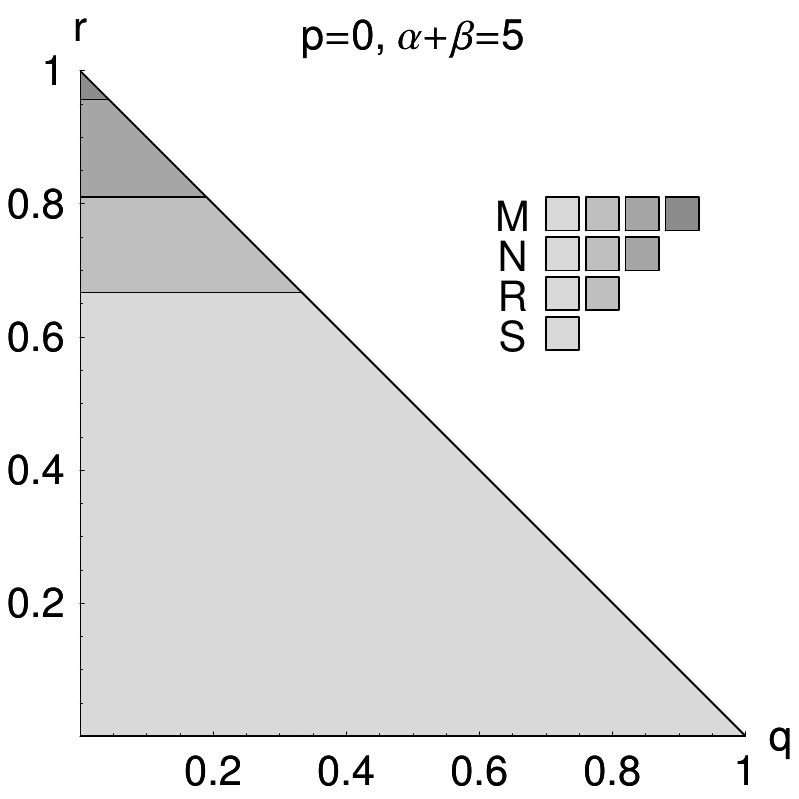}\includegraphics[width=0.3\textwidth]{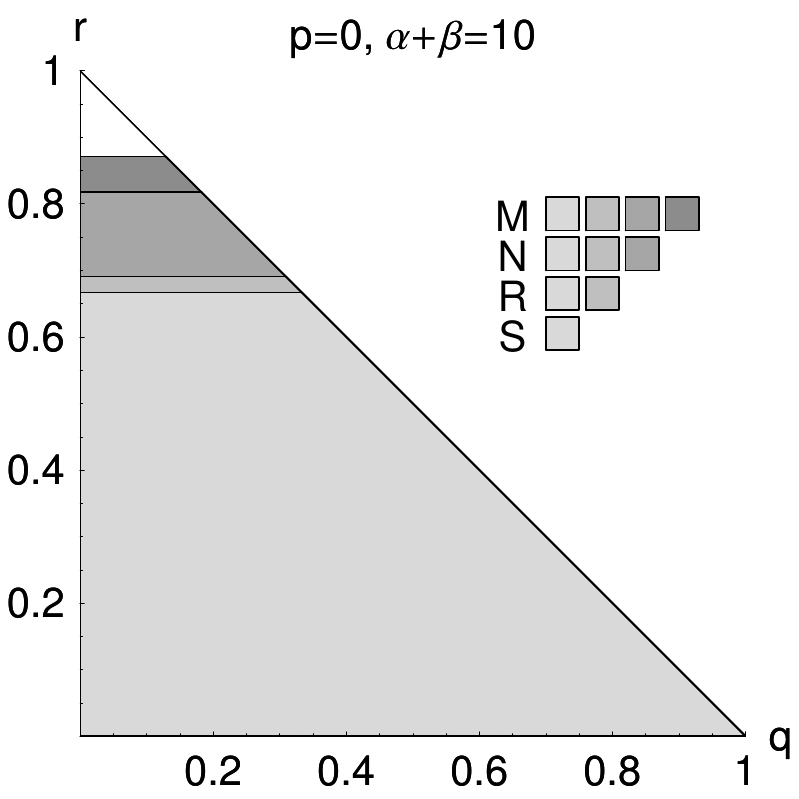}
\includegraphics[width=0.3\textwidth]{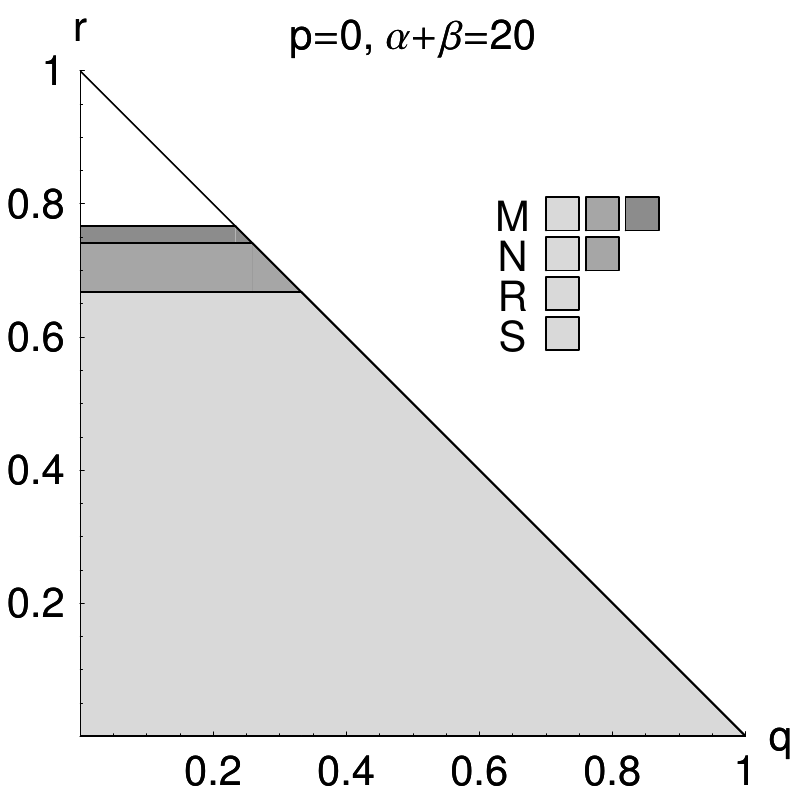}\\
\includegraphics[width=0.3\textwidth]{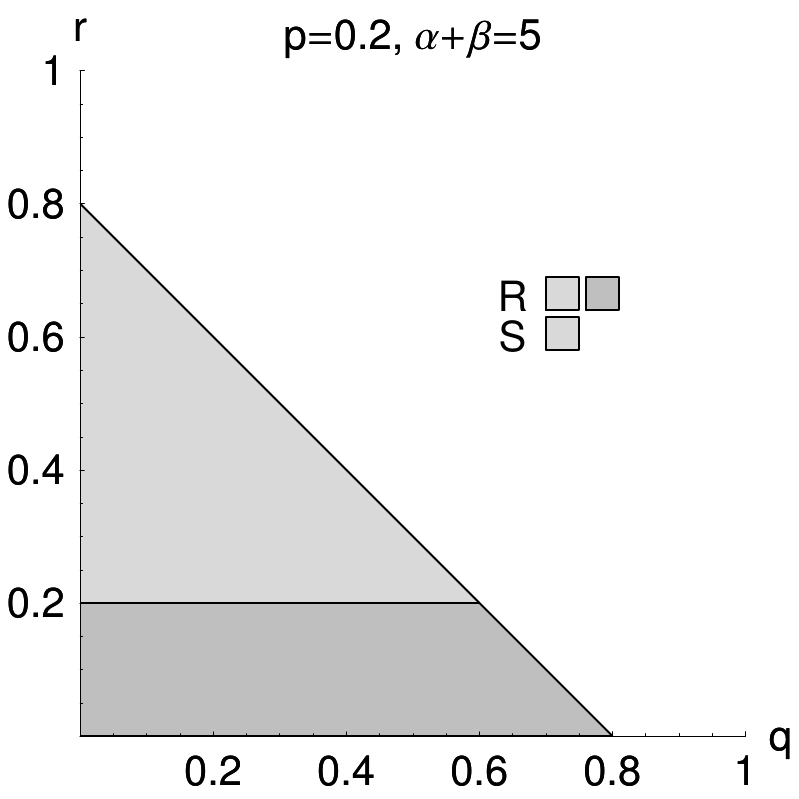}\includegraphics[width=0.3\textwidth]{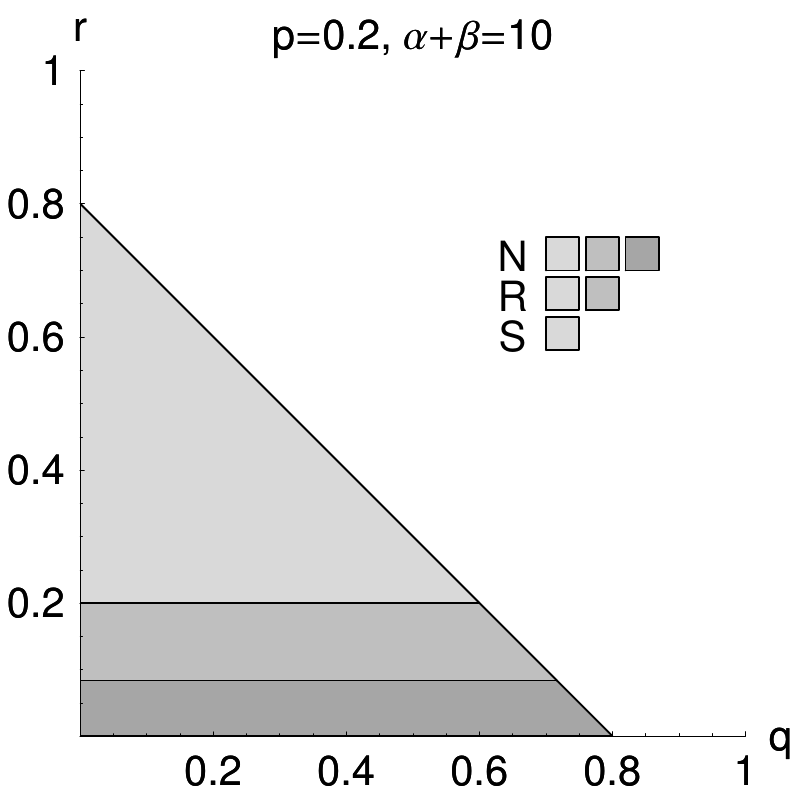}
\includegraphics[width=0.3\textwidth]{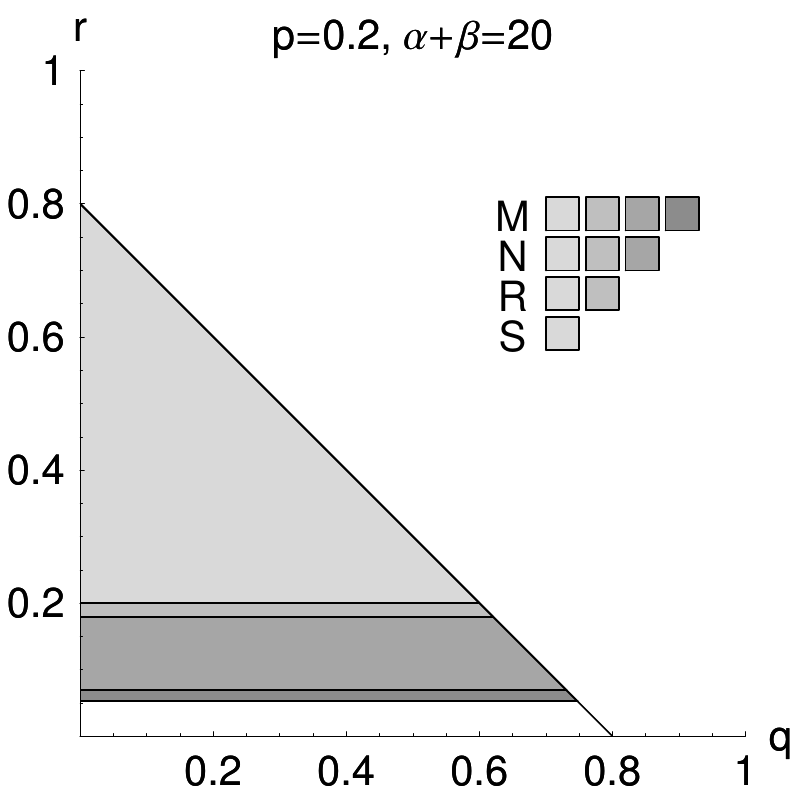}
  \caption{The sets of states (\ref{rot}) that fulfill the respective inequalities derived 
  from the Breuer map \cite{Breuer} (i.e. the map (15) taken with $U$ being a unitary antisymmetric matrix with the only nonzero entries $\pm 1$ lying on the antidiagonal) using the decomposition (\ref{PMdecomp})) are shown in different tones of gray. The set labeled M contains all states that satisfy inequality (\ref{nierownosc}), N - satisfy (\ref{TheoremIneq1}) with $\alpha\geq 1,\beta=1$, R - satisfy (\ref{TheoremIneq1}) with $\alpha=1,\beta\geq 1$, S - satisfy the Breuer--Hall criterion. If the state is not explicitly marked it means that it is the same as the largest set in the picture. Parameter $\alpha+\beta$ corresponds to the sum of powers; in the case of inequality (\ref{nierownosc})  we use the convention that $\beta=0$; parameter $p$ describes a considered class of states (see Eq. (\ref{rot})). The triangle marked in the figure denotes the set of all states.} \label{fig_nier_b}
\end{figure}

In this section, we show the effectiveness of derived inequalities in entanglement detection.
In the previous section, it was shown that the decomposition (\ref{PMdecomp}) can lead to inequalities 
involving entropies. We present the numerical results showing that inequalities arising from 
this decomposition detect more entanglement than those derived, e.g. from the minimal 
decomposition of a positive map. We apply the derived separability criterion to few classes of 
quantum states. First, the three parameter rotationally invariant $4\ot 4$ states
\begin{equation}\label{rot}
\varrho_{\mathrm{inv}}=p P_{0}+q P_{1}+r P_{2}+s P_{3},
\end{equation}
where $P_J$ is the projection onto the eigenspace of the square of total angular momentum 
$\mathbf{J}^2$ divided by the dimension of the corresponding eigenspace, i.e. $2J+1$. The total angular 
momentum takes values $J=|j_1-j_2|,\ldots,j_1+j_2$, where $j_1$ and $j_2$ are the local angular 
momenta, and $p,q,r$ and $s$ are nonnegative real numbers such that $p+q+r+s=1$. Both
subsystems of this states are maximally mixed. This class was extensively investigated, e.g. in \cite{Schliemann1,Schliemann2,BreuerSO1,BreuerSO2}. Second, a one parameter family of $4\otimes 4$ isotropic states
(which, actually, 
constitute a subset of rotationally invariant bipartite states presented above):
\begin{equation}\label{werner}
\varrho_{\mathrm{w}}=p
\frac{1}{4}\sum_{i,j=0}^{3}\ket{ii}\!\bra{jj}+(1-p)
\frac{\mathbbm{1}_{16}}{16}.
\end{equation}
The last analyzed class of states are the two--qubit states given in \cite{RH_PLA} and further analyzed 
in \cite{ZHHH}, i.e.
\begin{equation}\label{zh}
\varrho=q\proj{\Psi_1}+(1-q)\proj{\Psi_2},
\end{equation}
where $\ket{\Psi_1}=a\ket{00}+\sqrt{1-a^2}\ket{11}$ and $\ket{\Psi_2}=a\ket{10}+\sqrt{1-a^2}\ket{01}$, and the range of parameters is $0<a,q<1$.
\begin{figure}
\centering
\includegraphics[width=0.3\textwidth]{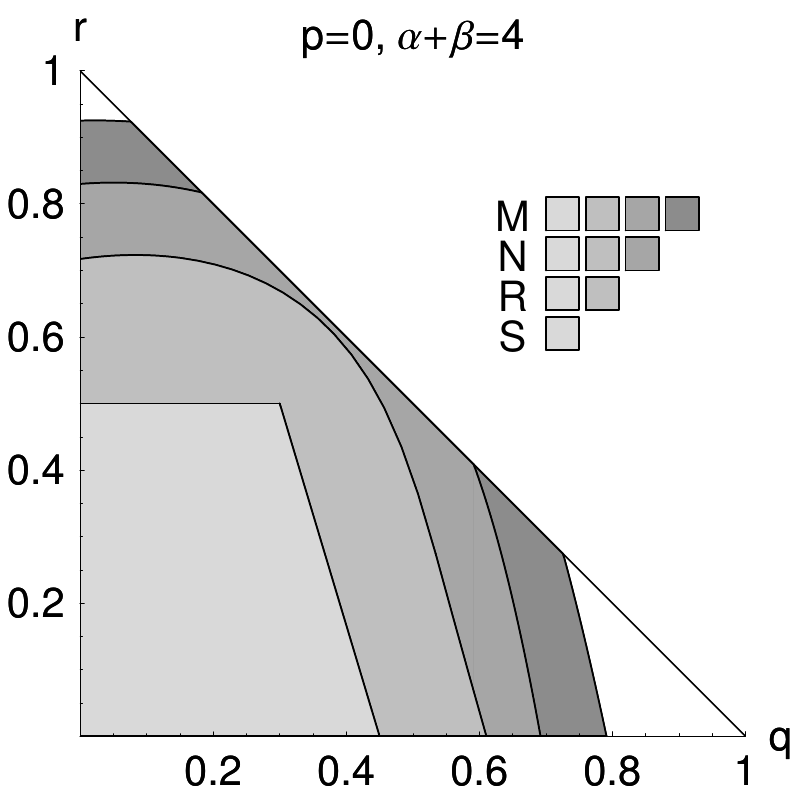}\includegraphics[width=0.3\textwidth]{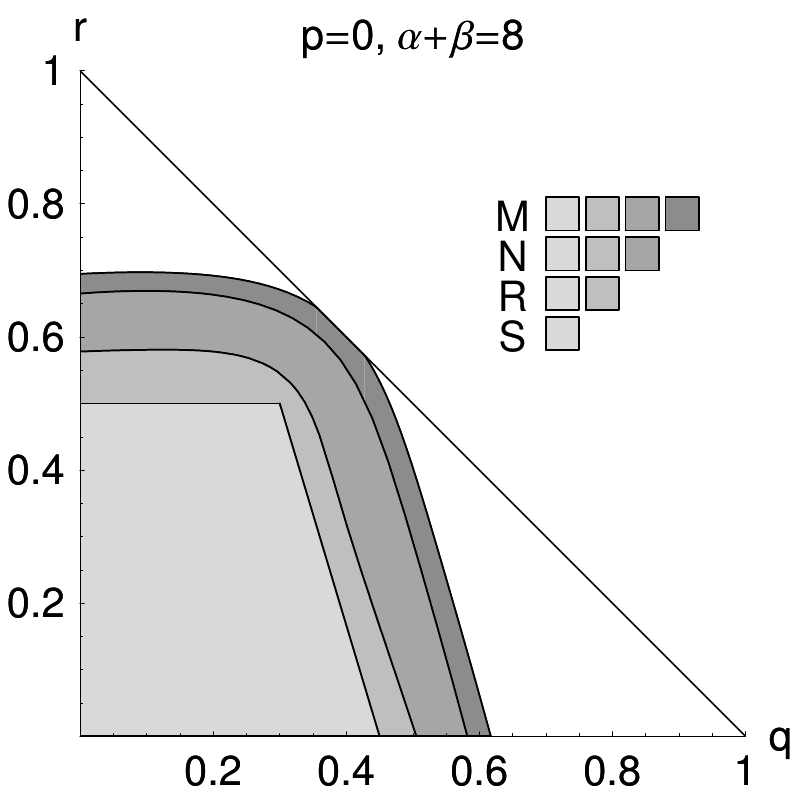}
\includegraphics[width=0.3\textwidth]{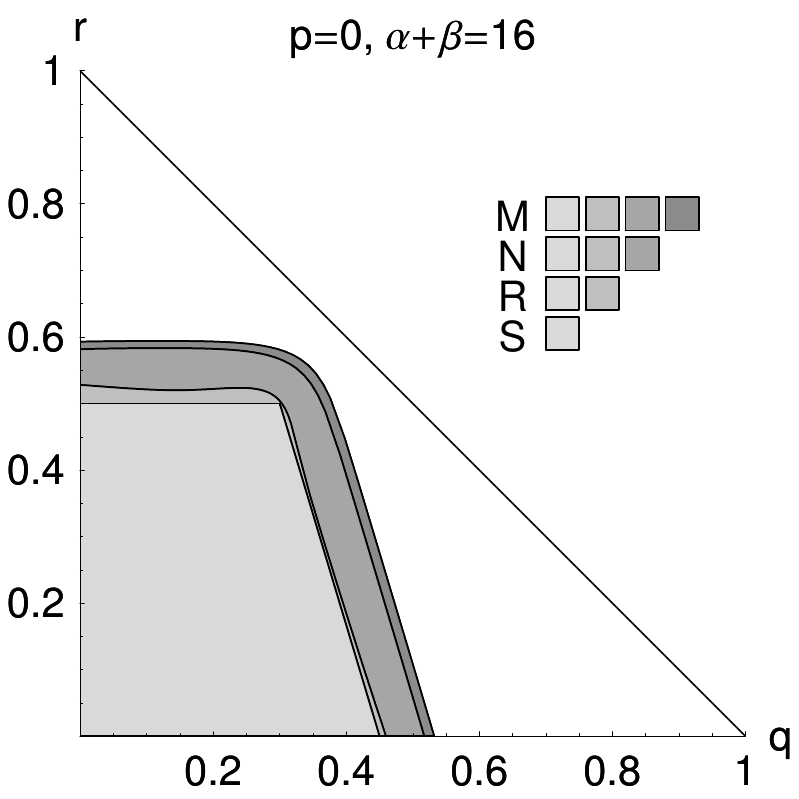}\\
\includegraphics[width=0.3\textwidth]{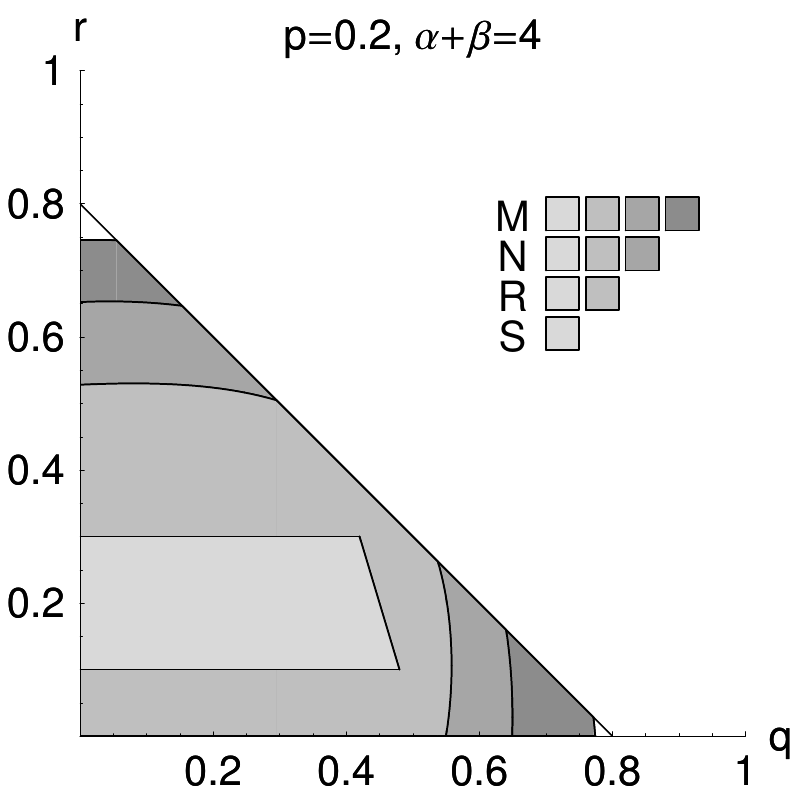}\includegraphics[width=0.3\textwidth]{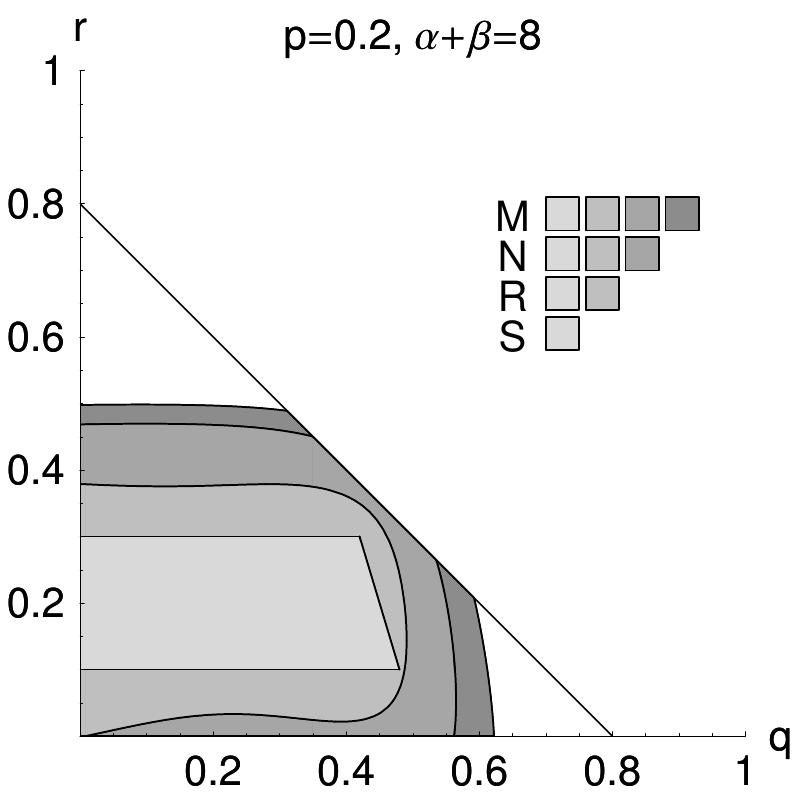}
\includegraphics[width=0.3\textwidth]{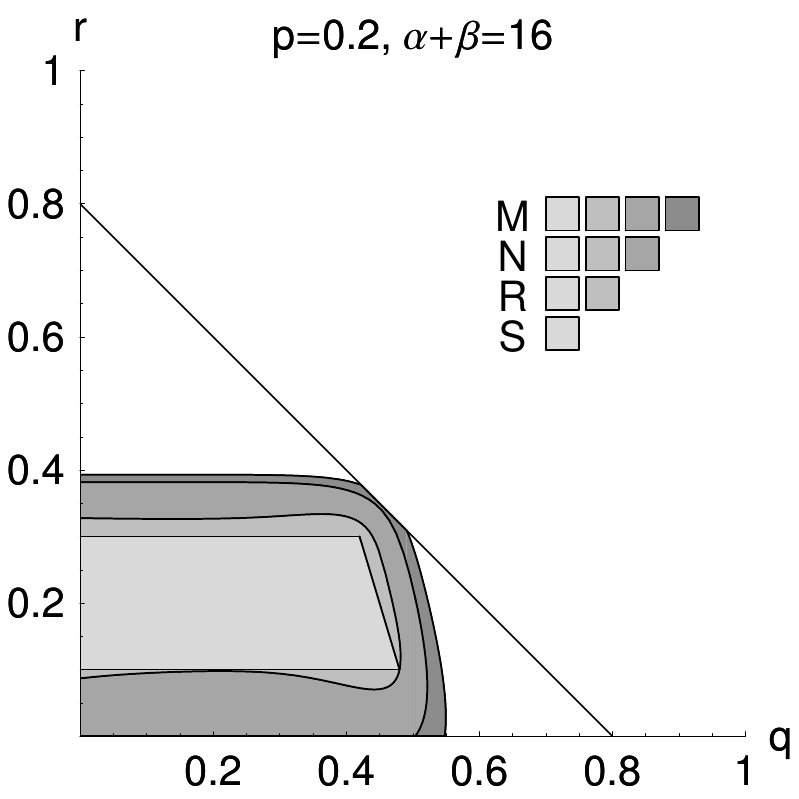}
  \caption{The sets of states (\ref{rot}) which fulfill the respective inequalities derived 
  from transposition map  (decomposition (\ref{PMdecomp})) are shown in different tones of 
  grey. The set labeled M contains all states that satisfy inequality (\ref{nierownosc}), N--
  satisfy (\ref{TheoremIneq1}) with $\alpha\geq 1,\beta=1$, R - satisfy (\ref{TheoremIneq1}) with $\alpha=1,\beta\geq 1$, S - satisfy the positive partial transposition criterion. Parameter $\alpha+\beta$ corresponds to the sum of powers; in the case of inequality (\ref{nierownosc}) we use the convention that $\beta=0$; parameter $p$ describes a
considered class of states (see Eq. (\ref{rot})). The triangle marked in the figure denotes the set of all states.}
\label{fig_nier_t}
\end{figure}
Let us first present the plots for inequalities of both types, i.e. of 
the form (\ref{nierownosc}) and (\ref{TheoremIneq1}) for rotationally invariant states (\ref{rot}). In Figure \ref{fig_nier_t} we compare the areas 
detected by the two inequalities arising from transposition map and taken with integer $\alpha \geq 1$. 
In Figure \ref{fig_nier_b} the same is shown for the Breuer-Hall map (with $U=V$, where $V$ 
is a unitary antisymmetric matrix with the only nonzero entries $\pm 1$ lying on the anti--diagonal).
Only the decomposition (\ref{PMdecomp}), corresponding to maximal length of $\Lambda_1$, is considered since it 
leads to inequalities for entropies similar to standard ones.

\begin{figure}
\centering
 \includegraphics[width=0.4\textwidth]{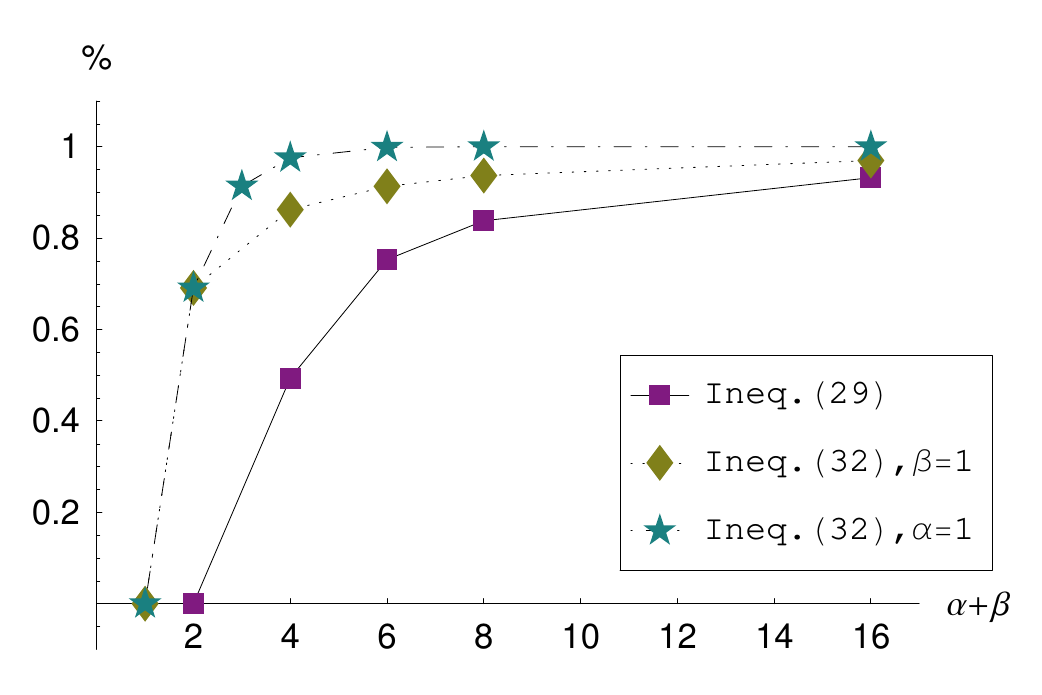}
  \caption{The percentage of the isotropic states (\ref{werner}) detected by the inequalities 
  derived from reduction map (decomposition (\ref{PMdecomp})) as a function of parameter 
  $\alpha+\beta$ ($\alpha+\beta$ corresponds to the sum of powers); in case of inequality (\ref{nierownosc}), $\beta=0$; parameter $p$ describes a considered class of states.} \label{fig_wern}
\end{figure}

As another example we show the effectiveness of inequalities derived from reduction 
map. We consider a class of isotropic states (\ref{werner}). In Figure \ref{fig_wern}, we show the percentage of
entangled states detected by inequalities presented here as a function of the total power $\alpha+\beta$ 
(corresponding to number of copies in a multicopy measurement). The percent is taken with 
respect to all states detected by the reduction map, and the measure applied hear is the Euclidean 
measure on the parameter space. All the measures were determined numerically. Note that in 
this case inequality (\ref{TheoremIneq1}) taken with $\beta=1$ is a standard entropic inequality (\ref{entrWithout1}).

Both provided examples confirm that for states with maximally mixed subsystem and 
particular choice of $\alpha+\beta$, the largest set of entangled states is detected by inequality (\ref{TheoremIneq1}) with 
$\alpha=1$ (the strongest criterion for states with maximally mixed subsystem), and the smallest by
inequality (\ref{nierownosc}) (the weakest from the separability criteria derived here). Moreover, the larger 
the power $\alpha+\beta$, the more states detect both inequalities. Analysis of states which do not have a 
maximally mixed subsystems (see below) will show that this is not a general rule.

\begin{figure}
\centering
  a)\includegraphics[width=0.4\textwidth]{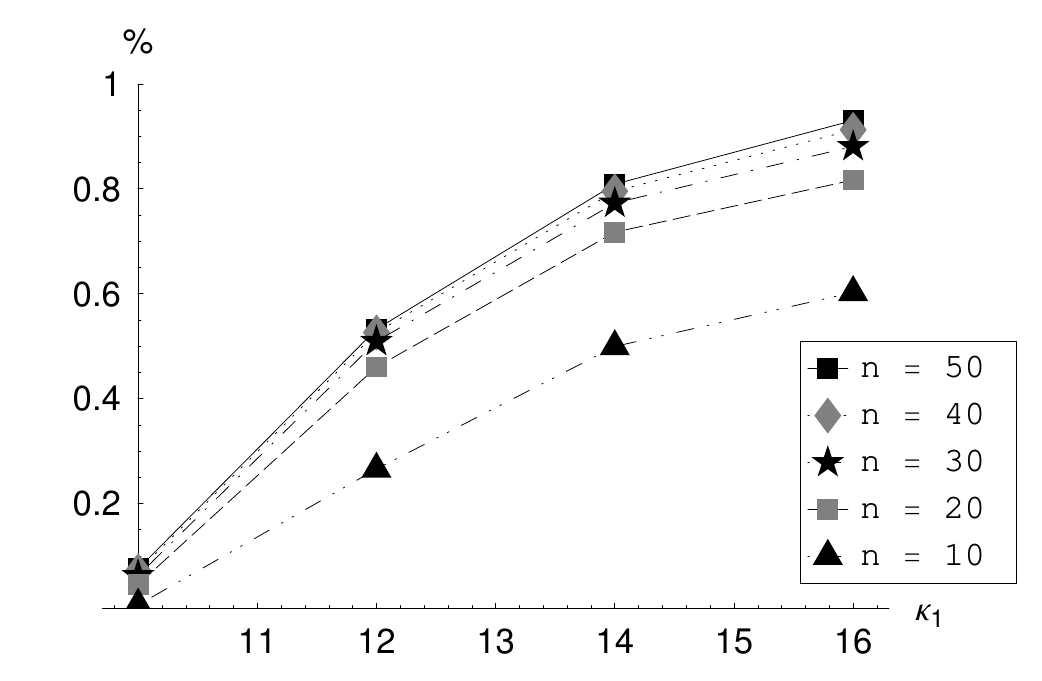}
  b)\includegraphics[width=0.4\textwidth]{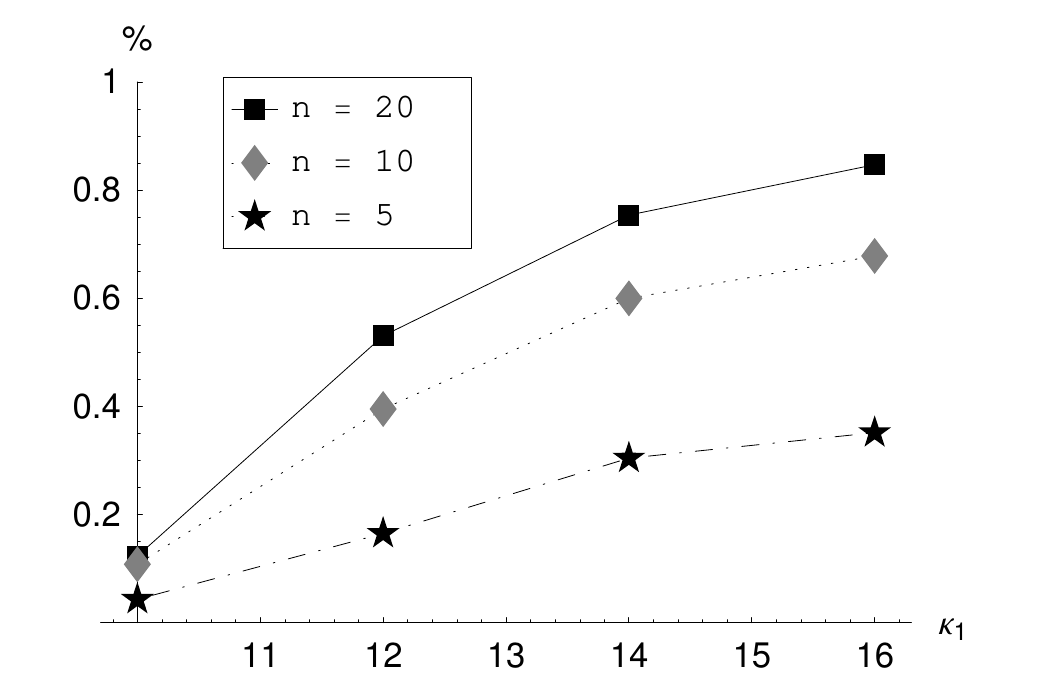}
  \caption{The percent of states (\ref{rot}) with $p=0$ detected by the inequalities 
  as a function of length of map $\Lambda_1$ in various decompositions of the transposition 
  map (described in the text). The plots were made for a) inequality (\ref{nierownosc}) and b) 
  inequality (\ref{TheoremIneq1}) with $\alpha>1,\beta=1$, $n=\alpha+\beta$. In each plot, 
  on the vertical axis the scale is $0$ to $1$, where $1$ corresponds to $100\%$.
  The percent of states detected by the inequalities is taken with respect to 
  the set of states detected by transposition map.}\label{fig_rrozklady1}
\end{figure}

It was already mentioned that the effectiveness of inequalities depends strongly on the 
choice of decomposition of a map. Let us now present this effect for inequalities derived from
transposition map and applied to rotationally invariant states (\ref{rot}) with $p=0$. The so--called 
minimal decomposition of the transposition map is given in Eq. (\ref{TranspDec}). The completely 
positive maps $T_{1(2)}$ in Eq. (\ref{TranspDec}) have the Kraus representation involving $SU(d)$ generators 
and identity. The map $T_1$ has minimal length $d(d+1)/2$ and $T_2$ has the minimal length 
$d(d-1)/2$. We obtain other decompositions of transposition map, with larger $\kappa_1$, by adding 
and subtracting the term $V_i\varrho V_i^{\dagger}$, which is in the Kraus representation of $T_2$ but not in the representation of $T_{1}$. In this way, the length of $\Lambda_2$ does not change (even though the map 
itself changed), and the length of $\Lambda_1$ is enlarged by one. Using this technique we got several 
different decompositions of the transposition map and checked how much entanglement in $4\otimes4$ 
rotationally invariant states (\ref{rot}) with $p=0$ can be detected by the inequalities with the same 
$\alpha$ but different $\Lambda_1$ and $\Lambda_2$. The results for inequality (\ref{nierownosc}) are presented in Figure \ref{fig_rrozklady1}a ($\kappa_1=16$ corresponds to decomposition from Table \ref{maps}, while $\kappa_1=10$ to equation (\ref{TranspDec})). The percent 
of detected entangled states is taken with respect to the amount of states detected by the 
transposition map itself, and the measure used here is the Euclidean measure on the parameter 
space which was determined numerically. The trend, similar for all values of parameter $\alpha$, is such 
that the larger the length of $\Lambda_1$ the more entanglement detects the respective inequality. A 
similar analysis conducted for inequality (\ref{TheoremIneq1}) presented in part (i) of the Theorem 2 for $\alpha\neq 1,\beta=1$ 
reveals a similar trend. The dependence on decomposition is shown in Fig. \ref{fig_rrozklady1}b. 

\begin{figure}
\centering
a)%
 \includegraphics[width=0.4\textwidth]{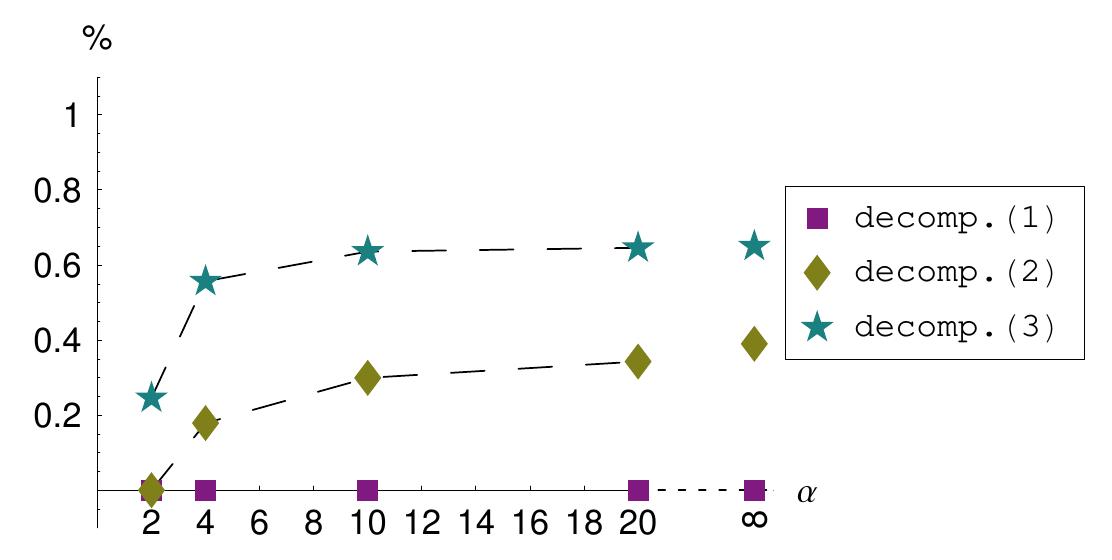}%
b)%
 \includegraphics[width=0.4\textwidth]{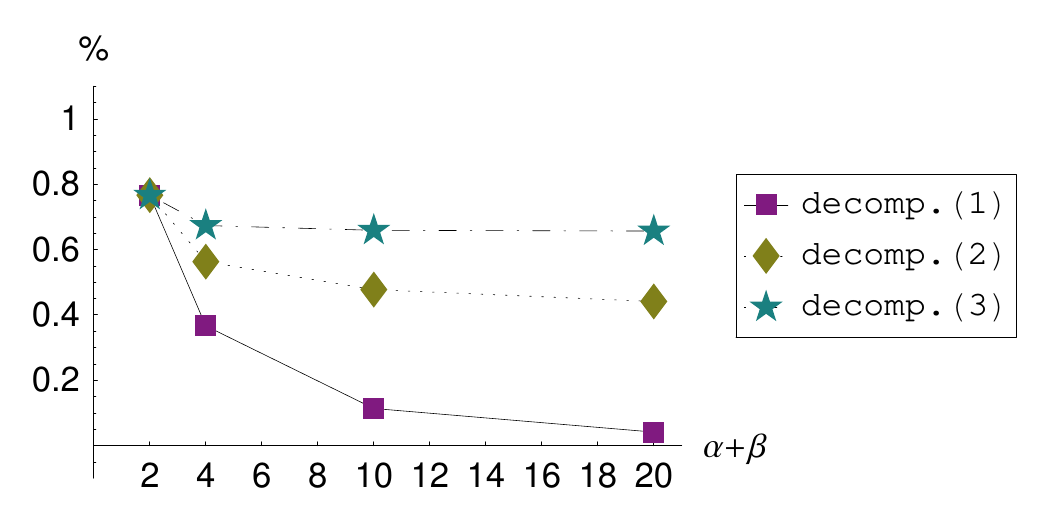}
  \caption{The percentage of states given by Eq. (\ref{zh}) detected by the inequalities derived from decompositions (1), (2) and (3) of the reduction map (defined in the text above the figure as $\Theta_i^{(1)}$, $\Theta_i^{(2)}$ and $\Theta_i^{(3)}$) acting on subsystem $A$, as a function of $\alpha$ or $\alpha+\beta$. The plots were made for a) inequality (\ref{nierownosc}) and b) inequality (\ref{TheoremIneq1}) with $\alpha>1,\beta=1$. In each plot, on the vertical axis the scale is $0$ to $1$, where $1$ corresponds to $100\%$.
  The percentage is taken with respect to all states (all entangled states are detected by reduction map).}
\label{fig_zha}
\end{figure}

\begin{figure}
\centering
a)%
 \includegraphics[width=0.4\textwidth]{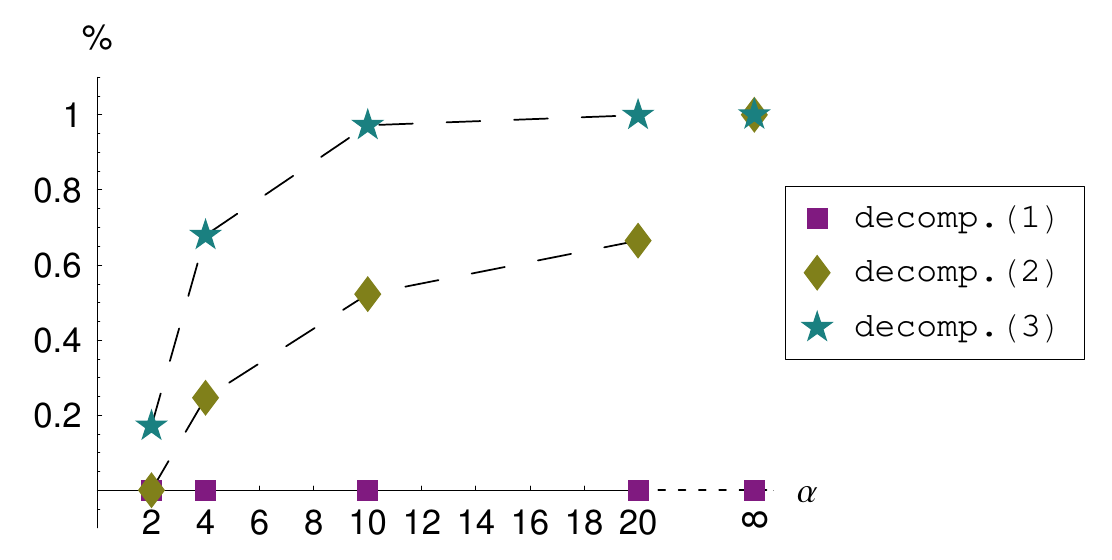}%
b)%
 \includegraphics[width=0.4\textwidth]{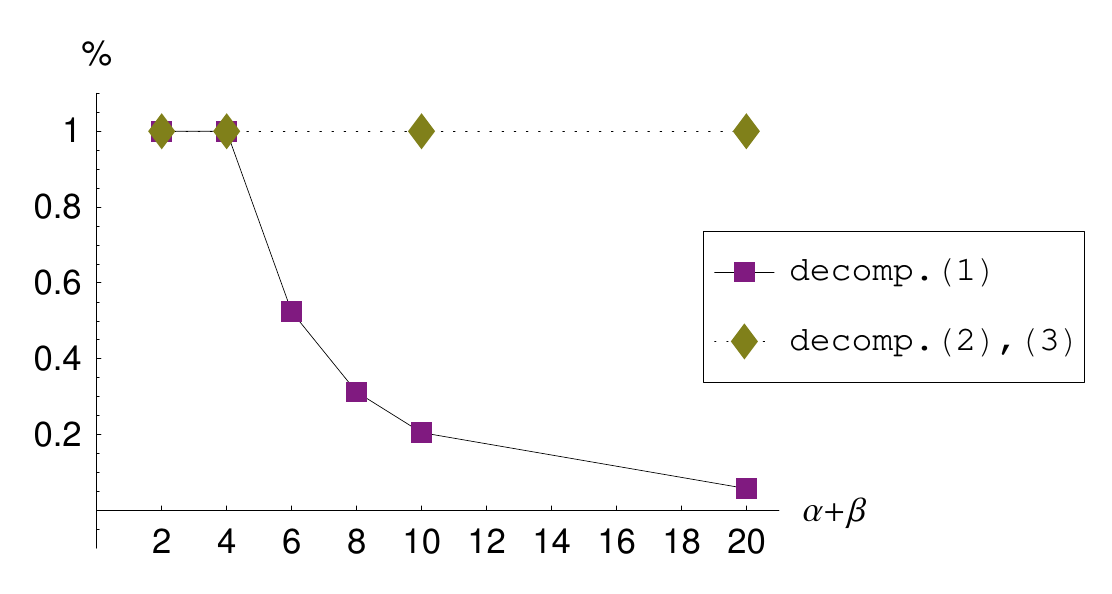}
  \caption{The percent of states given by Eq. (\ref{zh}) detected by the inequalities derived from various decompositions of the reduction map (described in the text) acting on subsystem $B$, versus parameters $\alpha$ and $\alpha+\beta$. The plots were made for a) inequality (\ref{nierownosc}), b) inequality (\ref{TheoremIneq1}) with $\alpha>1,\beta=1$.  In each plot, on the vertical axis the scale is $0$ to $1$, where $1$ corresponds to $100\%$.
  The percent is taken with respect to all states (all states are detected by reduction map).}
\label{fig_zhb}
\end{figure}

Let us now use the derived inequalities to analyze the separability of states (\ref{zh}), which do 
not have a maximally mixed subsystem, and are entangled for the whole range of parameters. 
We apply the inequalities derived from a few decompositions of the reduction criterion acting 
on subsystem $B$: firstly, the minimal decomposition, $\Theta_1^{(1)}(\varrho)=\varrho_A\ot
\mathbbm{1}_{d}-(1/2)\varrho$, $\Theta_2^{(1)}(\varrho)=(1/2)\varrho$, secondly, the decomposition with $\Lambda_{\Tr{}}$, $\Theta_1^{(2)}(\varrho)=\varrho_A\ot \mathbbm{1}_{d}$, $\Theta_2^{(2)}(\varrho)=\varrho$, 
and thirdly, $\Theta_1^{(3)}(\varrho)=\varrho_A\ot \mathbbm{1}_{d}+\varrho$, $\Theta_2^{(3)}(\varrho)=2\varrho$. Analogous decompositions are derived for the map acting on subsystem $A$. 
The results obtained are presented in Fig. \ref{fig_zha} for the map acting on 
subsystem $A$, and in Fig. \ref{fig_zhb} for subsystem $B$. In both figures the percent of entangled 
states detected by various inequalities is plotted versus parameter $\alpha+\beta$. Interestingly, for this 
class of states, increasing the parameter $\alpha+\beta$ does not always lead to a stronger separability 
criterion. When the inequality (\ref{TheoremIneq1}) with $\alpha\geq 1,\beta=1$ is considered, the larger parameter 
$\alpha$ the less entangled states is detected (see, e.g. Fig. \ref{fig_zha}b). Moreover, one can again see how 
the choice of decomposition influences effectiveness. In all figures, the greatest amount of 
detected entangled states corresponds to decomposition (3), whereas the smallest amount to 
minimal decomposition (1). Inequality (\ref{TheoremIneq1}) taken with $\alpha=1,\beta\geq 1$ does not depend on 
decomposition in this case since $\Lambda_2$ changes only up to a constant. Moreover, it detects all 
states for all $\beta\geq 1$.

\section{Conclusion}\label{conclusion_sec}

To summarize, firstly, we have presented a simple generalization of the Nielsen--Kempe disorder 
criterion to any possible positive map $\Lambda:M_{d}(\mathbb{C})\to M_{d}(\mathbb{C})$. However, the cost we have to pay 
for this generality is that instead of majorization we have to use the notion of weak majorization,
losing such a clear physical interpretation as in the case of majorization relations. Furthermore, 
our relations do not reproduce the Nielsen--Kempe result when the reduction map is considered. 
They only give equivalent criterion for states with at least one maximally mixed subsystem. 
On the other hand, taking other positive maps and in particular indecomposable ones, we can
obtain stronger separability criteria.

Still, the open question remains about possibility of derivation of majorization relations 
(not weak majorization) from any positive map, not only the reduction one. One of the possible 
ways is the following. Assume that $[I\ot\Lambda](\varrho)\geq  0$ with $\Lambda$ being some positive map for which equation (\ref{decompositionEta}) holds. Then rewrite the latter as 
\begin{equation}\label{conclusion1}
\varrho_{A}\ot \mathbbm{1}_{d}\geq \frac{\eta_{d}}{\xi}[I\ot
\Phi](\varrho ).
\end{equation}
Note that for the best known positive maps (see Table \ref{maps}) the above decomposition holds and 
$\eta_{d}\geq \xi$, which allows us to write
\begin{equation}
\varrho_{A}\ot \mathbbm{1}_{d}\geq [I\ot \Phi](\varrho ).
\end{equation}
Note, however, that above formula is a reduction criterion for the state $[I\ot\Phi](\varrho)$, which 
due to \cite{Hiroshima} leads straightforwardly to $\lambda(\varrho_{A})\succ\lambda([I\ot \Phi](\varrho ))$. This
relation, however, may be 
directly obtained from the Nielsen--Kempe majorization criterion, as for any quantum channel
and separable state $[I\ot\Phi](\varrho)$ is also separable state and must obey this criterion.

Secondly, we have provided two methods of deriving some entropic--like or entropic 
inequalities. The big advantage of the present approach is that, contrary to the method of 
\cite{RAJSPH,RAJS}, the present method does not require any assumptions about the investigated state, 
i.e. works for any bipartite state. The first of proposed method bases on the weak majorization 
criteria. To derive the second class of entropic inequalities, we utilized some class of functional 
inequalities. This is a continuation and extension of the results presented in \cite{RAJS} where we have
provided some inequalities stronger than the standard entropic inequalities and allowing for the 
detection of bound entanglement.

Moreover, it is pointed out that both the weak majorization criterion, and the inequalities 
derived from decomposition (\ref{PMdecomp}) of some map $\Lambda$ lead, for states with at least one maximally 
mixed subsystem, to the criterion equivalent to the necessary criterion $[I\ot \Lambda](\varrho)\geq 0$.

The derived generalizations of entropic inequalities were analyzed in the context of the 
protocol of state merging and approximation of a mean value of a linear entanglement witness. 
Moreover, all the derived inequalities (with integer $\alpha$ and $\beta$) contain expressions involving
products of operators, e.g. $\Tr[I\ot\Lambda_1(\varrho)]^{\alpha}[I\ot\Lambda_2(\varrho)]^{\beta}$, which can be (in principle) measured 
experimentally as a mean value of some operator (multi--copy entanglement witness) on a 
number of copies of a state. In the case of the first method (e.g. inequalities (\ref{nierownosc}), the number 
of copies is equal to $\alpha$, while in the case of inequalities (\ref{TheoremIneq1}), one has to take $\alpha+\beta$ copies of a given state at a time. A detailed analysis of this approach can be found in \cite{manycopy,RAJSPH}.

\section{Acknowledgments}
We gratefully acknowledge M. Horodecki and M. Lewenstein for discussions.
This work was prepared under the support of EU IP SCALA. R.A. gratefully acknowledges the support
from the Foundation for Polish Science and MEC (Spain) under the program Consolider-Ingenio 
2010 QOIT. J.S. acknowledges financial support from MCI (Spain) under the contract FIS2008-01236/FIS and the `Universitat Aut\`{o}noma de Barcelona'.


\section*{References}
\begin{thebibliography}{10}
\bibitem{przegladowka} R. Horodecki, P. Horodecki, M. Horodecki, K. Horodecki, Rev. Mod. Phys. {\bf 81}, 865 (2009).

\bibitem{Werner}R. F. Werner, Phys. Rev. A {\bf 40}, 4277 (1989).

\bibitem{Peres}A. Peres, Phys. Rev. Lett. {\bf 77}, 1413 (1996).

\bibitem{HHH_entr}R. Horodecki, M. Horodecki, and P. Horodecki,
Phys. Lett. A {\bf 210}, 227 (1996).

\bibitem{mapy}M. Horodecki, P. Horodecki, and R. Horodecki,
Phys. Lett. A {\bf 223}, 1 (1996).

\bibitem{range}P. Horodecki, Phys. Lett. A {\bf 232}, 333 (1997).

\bibitem{NielsenKempe}M. A. Nielsen and J. Kempe, Phys. Rev. Lett. {\bf 86}, 5148 (2001).

\bibitem{Korbicz1}J. K. Korbicz, J. I. Cirac, and M. Lewenstein, Phys. Rev. Lett. {\bf 95}, 120502 (2005); 
Phys. Rev. Lett. {\bf 95}, 259901 (2005).

\bibitem{Korbicz}J. K. Korbicz, O. G\"{u}hne, M. Lewenstein, H. H\"{a}ffner, C. F. Roos, and R. Blatt, Phys. Rev. A {\bf 74}, 052319 (2006).

\bibitem{Guhne}O. G\"{u}hne and N. L\"{u}tkenhaus, Phys. Rev. Lett. {\bf 96}, 170502 (2006).

\bibitem{deVicente}J. de Vicente, Quantum Inf. Comput. {\bf 7}, 624 (2007).

\bibitem{ChengJieZhang} C.--J. Zhang, Y.-S. Zhang, S. Zhang, G.-C. Guo, Phys. Rev. A {\bf 77}, 060301(R) (2008).

\bibitem{covariance} O. G\"{u}hne, P. Hyllus, O. Gittsovich, J. Eisert, Phys. Rev. Lett. {\bf 99}, 130504 (2007).

\bibitem{SperlingVogel} J. Sperling and W. Vogel, Phys. Rev. A {\bf 79}, 022318 (2009).

\bibitem{SpinSq} G. T\'oth, C. Knapp, O. G\"{u}hne, and H. J. Briegel, Phys. Rev. Lett. {\bf 99}, 250405 (2007);
G. T\'oth, C. Knapp, O. G\"{u}hne, and H. J. Briegel, Phys. Rev. A. {\bf 79}, 042334 (2009).

\bibitem{EntDet} O. G\"{u}hne and G. T\'oth, Phys. Rep. {\bf 474}, 1 (2009).

\bibitem{pure} S. P. Walborn, P. H. Souto Ribeiro, L. Davidovich, F. Mintert, and A. Buchleitner, Nature {\bf 440}, 1022 (2006); L. Aolita and F. Mintert, Phys. Rev. Lett. {\bf 97}, 050501 (2006).

\bibitem{RAMDPH} R. Augusiak, M. Demianowicz, and P. Horodecki, Phys. Rev. A {\bf 77}, 030301(R) (2008).

\bibitem{reduction} M. Horodecki and P. Horodecki, Phys. Rev. A {\bf 59}, 4206 (1999).

\bibitem{CerfAdamiGingrich} N. J. Cerf, C. Adami, and R. M. Gingrich, Phys. Rev. A {\bf 60}, 898 (1999).

\bibitem{maps_exp}P. Horodecki and A. Ekert, Phys. Rev. Lett. {\bf 89}, 127902 (2002);
P. Horodecki, Phys. Rev. Lett. {\bf 90}, 167901 (2003); 
H. A. Carteret, Phys. Rev. Lett. {\bf 94}, 040502 (2005);
P. Horodecki, R. Augusiak, and M. Demianowicz, Phys. Rev. A {\bf 74}, 052323 (2006).

\bibitem{RHPH}R. Horodecki and P. Horodecki, Phys. Lett. A {\bf 194}, 147 (1994).

\bibitem{MHJOAW} M. Horodecki, J. Oppenheim, A. Winter, Nature {\bf 436}, 673 (2005); 
M. Horodecki, J. Oppenheim, A. Winter, Commun. Math. Phys. {\bf 269}, 107 (2007).

\bibitem{RHMH}R. Horodecki and M. Horodecki, Phys. Rev. A {\bf 54}, 1838 (1996).

\bibitem{Terhal}B. M. Terhal, Theor. Comput. Sci. {\bf 287}, 313 (2002).

\bibitem{VollbrechtWolf}K. G. H. Vollbrecht and M. Wolf, J. Math. Phys. {\bf 43}, 4299 (2002).

\bibitem{Renyi} A. R\'{e}nyi, Proc. Fourth Berkeley Symp. on Math. Statist. and Prob., Vol. {\bf 1} (Univ. of Calif. Press, 1961).

\bibitem{Tsallis} C. Tsallis, J. Stat. Phys. {\bf 52}, 479 (1988).

\bibitem{Wehrl}A. Wehrl, Rev. Mod. Phys. {\bf 50}, 221 (1978).

\bibitem{Investigation}S. Abe and A. K. Rajagopal, Physica A {\bf 289}, 157 (2001);
C. Tsallis, S. Lloyd, and M. Baranger, Phys. Rev. A {\bf 63}, 042104 (2001);
J. Batle, M. Casas, A. Plastino, and A. R. Plastino, {\it ibid.} {\bf 71}, 024301 (2005), and references therein.

\bibitem{ZHHH} K. \.{Z}yczkowski, P. Horodecki, M. Horodecki, R. Horodecki, Phys. Rev. A {\bf 65} 012101 (2002).

\bibitem{RossignoliCanosa} R. Rossignoli and N. Canosa, Phys. Rev. A {\bf 66}, 042306 (2002);
R. Rossignoli and N. Canosa, {\it ibid.} {\bf 67}, 042302 (2003).

\bibitem{manycopy} P. Horodecki, Phys. Rev. A {\bf 68}, 052101 (2003).

\bibitem{Bovino} F. A. Bovino, G. Castagnoli, A. Ekert, P. Horodecki, C. Moura Alves, and A. V. Sergienko, Phys. Rev. Lett. {\bf 95}, 240407 (2005).
    
\bibitem{Schmid}Ch. Schmid, N. Kiesel, W. Wieczorek, H. Weinfurter, F. Mintert, and A. Buchleitner, Phys. Rev. Lett. {\bf 101}, 260505 (2008).
    
\bibitem{Jaksch} C. Moura Alves and D. Jaksch, Phys. Rev. Lett {\bf 93}, 110501 (2004).

\bibitem{Hiroshima} T. Hiroshima, Phys. Rev. Lett. {\bf 91}, 057902 (2003).

\bibitem{BE} M. Horodecki, P. Horodecki, and R. Horodecki, Phys. Rev. Lett. {\bf 80}, 5239 (1998).

\bibitem{RAJSPH}R. Augusiak, J. Stasi\'nska, and P. Horodecki, Phys. Rev. A {\bf 77}, 012333 (2008).

\bibitem{RAJS}R. Augusiak and J. Stasi\'nska, Phys. Rev. A {\bf 77}, 010303(R) (2008).

\bibitem{AlbertiUhlmann} P. M. Alberti and A. Uhlmann, {\it Stochasticity and Partial Order---Doubly Stochastic Maps and Unitary Mixing}, Mathematics and its Applications vol. {\bf 9} (D.Reidel Publ. Company, Dordrecht, 1982).

\bibitem{NielsenConditions} M. A. Nielsen, Phys. Rev. Lett. {\bf 83}, 436 (1999).

\bibitem{NielsenVidal}M. A. Nielsen and G. Vidal, Quantum Inf. Comput. {\bf 1}, 76 (2001).

\bibitem{Bhatia}R. Bhatia, {\it Matrix Analysis} (Springer, New York, 1997).

\bibitem{Marshall} A. W. Marshall and I Olkin, {\it Inequalities: Theory of Majorization and Its Applications}, Mathematics in Sciences and Engineering Vol. {\bf 143}, (Academic Press, New York, New York, 1979).

\bibitem{Jam} A. Jamio\l{}kowski, Rep. Math. Phys. {\bf 3}, 275 (1972).

\bibitem{Choi_iso} M.-D. Choi, Linear Alg. Appl. {\bf 10}, 285 (1975).

\bibitem{Hou}J.-C. Hou, J. Oper. Theory {\bf 39}, 43 (1998).

\bibitem{postacKrausa}K. Kraus, States, {\it Effects and Operations: Fundamental Notions of Quantum Theory},
(Wiley, New York, 1991).

\bibitem{Jamiolkowski1}A. Jamio\l{}kowski, Open Sys. and Inf. Dyn. {\bf 11}, 385-390 (2004).

\bibitem{Breuer}H.--P. Breuer, Phys. Rev. Lett. {\bf  97}, 080501 (2006).

\bibitem{Hall}W. Hall, J. Phys. A: Math. Theor. {\bf 40}, 6183 (2007).

\bibitem{Arimoto}S. Arimoto, Inf. Control {\bf 19}, 181 (1971).

\bibitem{Lowner}K. L\"{o}wner, Math. Z. {\bf 38}, 177 (1934).

\bibitem{Lindblad}G. Lindblad, Commun. Math. Phys. {\bf 28}, 245 (1972).

\bibitem{WernerHolevo} R. F. Werner, A. S. Holevo, J. Math. Phys. {\bf 43}, 4353 (2002).

\bibitem{Choi}M.-D. Choi, Linear Algebr. Appl. {\bf 12}, 95 (1975).

\bibitem{uogolnieniaChoi}K. Tanahasi and J. Tomiyama, Can. Math. Bull. {\bf 31}, 308 (1988); 
H. Osaka, Linear Algebr. Appl. {\bf 153}, 73 (1991); H. Osaka {\it ibid} {\bf 186}, 45 (1993).

\bibitem{Ha1}K.-C. Ha, Publ. Res. Inst. Math. Sci. {\bf 34}, 591 (1998).

\bibitem{Schliemann1}J. Schliemann, Phys. Rev. A {\bf 68 }, 012309 (2003).

\bibitem{Schliemann2}J. Schliemann, Phys. Rev. A {\bf 72}, 012307 (2005).

\bibitem{BreuerSO1}H.--P. Breuer, Phys. Rev. A {\bf 71}, 062330 (2005).

\bibitem{BreuerSO2}H.--P. Breuer, J. Phys. A {\bf 38}, 9019 (2005).

\bibitem{RH_PLA} R. Horodecki, Phys. Lett. A {\bf 210}, 223 (1996).

\end{thebibliography}
\end{document}